\documentclass[iop, apj, twocolapendix]{emulateapj} 
\usepackage{lmodern}
\usepackage{amssymb,amsmath}
\usepackage[T1]{fontenc}
\usepackage[utf8x]{inputenc}
\usepackage{aas_macros}
\usepackage{natbib}
\usepackage{graphicx}
\usepackage[unicode=true]{hyperref}
\hypersetup{breaklinks=true,
            bookmarks=true,
            pdfauthor={},
            pdftitle={The Lyman alpha Reference Sample: V. The impact of
	    neutral ISM kinematics and geometry on Lyman alpha escape},
            colorlinks=true,
            citecolor=blue,
            urlcolor=blue,
            linkcolor=magenta,
            pdfborder={0 0 0}}
\shorttitle{LARS V: HST-COS UV Spectroscopy}
\shortauthors{T. E. Rivera-Thorsen et al.}
\slugcomment{Accepted for publication in \apj}

\begin{document}

\title{\textbf{The Lyman alpha Reference Sample: V. The impact of
       neutral ISM kinematics and geometry on Lyman alpha escape
       \footnotemark[\(\ast\)]}}
\footnotetext[\(\ast\)]{Based on observations made with the NASA/ESA Hubble Space Telescope,
obtained at the Space Telescope Science Institute, which is operated by
the Association of Universities for Research in Astronomy, Inc., under
NASA contract NAS 5-26555. These observations are associated with
programs GO11522, GO11727, GO12027, and GO12583.}

\author{Thøger E. Rivera-Thorsen\altaffilmark{1, $\star$}}
\author{Matthew Hayes\altaffilmark{1}}
\author{Göran Östlin\altaffilmark{1}}
\author{Florent Duval\altaffilmark{1}}
\author{Ivana Orlitová\altaffilmark{2}}
\author{Anne Verhamme\altaffilmark{3}}
\author{J. Miguel Mas-Hesse\altaffilmark{4}}
\author{Daniel Schaerer\altaffilmark{3, 5}}
\author{John M. Cannon\altaffilmark{6}}
\author{Héctor Otí-Floranes\altaffilmark{7, 8}}
\author{Andreas Sandberg\altaffilmark{1}}
\author{Lucia Guaita\altaffilmark{1, 13}}
\author{Angela Adamo\altaffilmark{1}}
\author{Hakim Atek\altaffilmark{11}}
\author{E. Christian Herenz\altaffilmark{9}}
\author{Daniel Kunth\altaffilmark{10}}
\author{Peter Laursen\altaffilmark{12}}
\author{Jens Melinder\altaffilmark{1}}

\altaffiltext{$\star$}{For correspondence regarding this article, please
write to T.~E. Rivera-Thorsen \email{trive@astro.su.se}}
\altaffiltext{1}{Department of Astronomy, Oskar Klein Centre, Stockholm University,
AlbaNova University Centre, SE-106 91 Stockholm, Sweden}
\altaffiltext{2}{Astronomical Institute, Academy of Sciences of the Czech Republic, Boční
II, CZ-14131 Prague, Czech Republic}
\altaffiltext{3}{Geneva Observatory, University of Geneva, 51 Chemin des Maillettes,
CH-1290 Versoix, Switzerland}
\altaffiltext{4}{Centro de Astrobiología (CSIC--INTA), Departamento de Astrofísica, POB
78, E--28691 Villanueva de la Ca\~nada, Spain}
\altaffiltext{5}{Université de Toulouse; UPS-OMP; IRAP; Toulouse, France}
\altaffiltext{6}{Department of Physics and Astronomy, Macalester College, 1600 Grand
Avenue, Saint Paul, MN 55105, USA}
\altaffiltext{7}{Instituto de Astronomía, Universidad Nacional Autónoma de México, Apdo.
Postal 106, B. C. 22800 Ensenada, Mexico}
\altaffiltext{8}{Centro de Radioastronomía y Astrofísica, UNAM, Campus
Morelia, Michoac\'an, C.P. 58089, Mexico}
\altaffiltext{9}{Leibniz-Institut für Astrophysik (AIP), An der Sternwarte 16, D-14482
Potsdam, Germany.A}
\altaffiltext{10}{Institut d'Astrophysique de Paris, UMR 7095 CNRS \& UPMC, 98 bis Bd
Arago, 75014 Paris, France.}
\altaffiltext{11}{Laboratoire d'Astrophysique, École Polytechnique Fédérale de Lausanne
(EPFL), Observatoire, CH-1290 Sauverny, Switzerland.}
\altaffiltext{12}{Dark Cosmology Centre, Niels Bohr Institute, University of Copenhagen,
Juliane Maries Vej 30, 2100 Copenhagen, Denmark}
\altaffiltext{13}{INAF - Osservatorio Astronomico di Roma, Via Frascati 33,
I–00040, Monteporzio, Italy}

\begin{abstract} 

We present high-resolution far-UV spectroscopy of the 14 galaxies of the
Lyman Alpha Reference Sample; a sample of strongly star-forming galaxies at low
redshifts ($0.028 < z < 0.18$). We compare the derived properties to global
properties derived from multi band imaging and 21 cm HI interferometry and
single dish observations, as well as archival optical SDSS spectra. Besides the
Lyman $\alpha$ line, the spectra contain a number of metal absorption features
allowing us to probe the kinematics of the neutral ISM and evaluate the optical
depth and and covering fraction of the neutral medium as a function of
line-of-sight velocity.  Furthermore, we show how this, in combination with
precise determination of systemic velocity and good Ly$\alpha$ spectra, can be
used to distinguish a model in which separate clumps together fully cover the
background source, from the ``picket fence'' model named by
\citet{Heckman2011}.  We find that no one single effect dominates in governing
Ly$\alpha$ radiative transfer and escape.  Ly$\alpha$ escape in our sample
coincides with a maximum velocity-binned covering fraction of $\lesssim 0.9$ and
bulk outflow velocities of $\gtrsim 50$ km s$^{-1}$, although a number of
galaxies show these characteristics and yet little or no Ly$\alpha$ escape. We
find that Ly$\alpha$ peak velocities, where available, are not consistent with a
strong backscattered component, but rather with a simpler model of an intrinsic
emission line overlaid by a blueshifted absorption profile from the outflowing
wind. Finally, we find a strong anticorrelation between H$\alpha$ equivalent
width and maximum velocity-binned covering factor, and propose a heuristic
explanatory model.  

%  We here present an analysis of high-resolution ultraviolet spectra of the 14
%  galaxies of the LARS sample obtained with the Cosmic Origins Spectrograph at
%  the Hubble Space Telescope, comparisons with global properties of the same
%  galaxies inferred from UV and optical HST imaging. We discuss the
%  implications for radiation transportation and the properties of star-forming
%  galaxies in general. We find that Ly\(\alpha\) escape in our sample coincides
%  with a maximum velocity-binned covering fraction of \(\lesssim 0.9\) and an
%  outflow wind velocity \(\gtrsim 50\, \mathrm{km\, s}^{-1}\).  However, a
%  number of galaxies in the sample show significant outflow velocities but
%  little or no Ly\(\alpha\) escape, indicating that outflows, while necessary,
%  are not sufficient to allow Ly\(\alpha\) escape.  Furthermore, we find a
%  clear anticorrelation between H\(\alpha\) equivalent width and maximum
%  velocity-binned covering fraction. We propose an interpretation which further
%  develops the model presented by \citet{TenorioTagle99} and elaborated by
%  \citet{MasHesse03}, in which Rayleigh-Taylor instabilities in an expanding
%  spherical H {\sc i} shell, induced by star formation feedback, lead to
%  breakup and clumping of the neutral medium in clouds of various velocities,
%  embedded in a surrounding ionized medium.  
\end{abstract}

\section{Introduction}\label{sec:intro}

Radiation in Lyman alpha (Ly\(\alpha\)), the transition between the
ground state and the first excited state of atomic Hydrogen, is one of
the most important sources of information about the high-redshift
Universe; yet interpretation of this line has proven complicated.

Ly\(\alpha\) reprocesses around 2/3 of ionizing photons, or 1/3 of raw
ionizing power, \citep{Spitzer1978, DijkstraRev} from star-forming regions
into one strong, narrow emission feature. The transition is resonant which,
combined with typically high column densities of neutral Hydrogen in young
galaxies, leads to a high probability of absorption by ground-state hydrogen
atoms and immediate reemission in a different direction. As a result, the
typical mean free path of a Ly\(\alpha\) photon is considerably shorter and
the optical path before escape significantly longer than is the case for
non-resonance lines. Due to the longer path traveled within a H {\sc i}
system, Ly\(\alpha\) photons have a larger probability of encountering and
getting scattered or absorbed by dust grains.

While the gas will be Doppler shifted out of resonance with a fraction
of the Ly\(\alpha\) photons which can then escape freely, the photons
that do experience scattering will typically be transported far away
from their projected spatial point of origin after a number of
scatterings before escaping or being absorbed. This effectively erases
any information about their spatial origin, making Ly\(\alpha\) appear
as an extended, diffuse halo superimposed upon imaging data in other
filters \citep{Ostlin2009, LARS0, LARSII, Steidel2011}. With the
escaping radiation distributed over a large area, these halos tend to
contain large regions of low surface brightness which may integrate to a
significant amount of the escaping Ly\(\alpha\) flux which is difficult
to detect against foreground and noise, especially at high \(z\). The
strength of this effect varies strongly between galaxies; some show
strong diffusion, while it is less pronounced in others. Observed
brightness and morphology in Ly\(\alpha\) is thus in part decoupled from
intrinsic brightness and morphology, with observed values only in the
broadest statistical sense tracing the intrinsic configuration.

The strong resonance of the transition has a similar impact on the shape
and strength of the spectroscopic signature of Ly\(\alpha\). As shown by
e.g. \citet{Neufeldt1991}, it is very unlikely, even with a very low
dust content, for Ly\(\alpha\) to escape a static, homogeneous H\textsc{i}
medium. However, as observed by \citet{Kunth98} and modeled by e.g.
\citet{Verhamme2006}, \citet{Dijkstra2006}, \citet{Laursen2009b}, the
presence of a bulk outflow, generated by e.g.~radiation or momentum
pressure from star-forming regions, can Doppler shift the atoms of the
outflowing gas out of resonance with the Ly\(\alpha\) radiation coming
from the inner regions, allowing them to escape more easily. Galaxies of
similar intrinsic Ly\(\alpha\) equivalent width may show observed
spectral features ranging from strong, P Cygni-like features to broad,
damped absorption, depending on e.g.~outflow velocity. Even in the
presence of outflows, dust still has a suppressing impact on
Ly\(\alpha\) escape fraction \citep{Atek2009, LARS0}. Studies have also
suggested that observed equivalent widths of Ly\(\alpha\) are better
reproduced by a clumpy medium embedded in a warm or hot, diffuse medium
than the earlier models of a homogeneous expanding shell
\citep{Laursen2013, Duval2014}. While many constraints and conclusions
have been reached, pinning down the exact interplay of factors which
governs Ly\(\alpha\) radiative transfer processes has proven difficult.
The importance of outflows has been observationally confirmed by e.g.
\citet{Kunth98} and \citet{Wofford2013}. \citet{MasHesse03} conclude
that also the spatial distribution of neutral gas along the line of
sight is important to the shape of the line profile, and have proposed a
scenario reconciling the models with general theory of galaxy formation.

The Lyman Alpha Reference Sample (LARS, \citet{LARS0}) is a coordinated
attempt to resolve some of the outstanding questions of Ly\(\alpha\)
radiative transfer and escape by careful and detailed study of a sample
of 14 star-forming galaxies in the nearby Universe. The backbone of the
study is made up of HST imaging in H\(\alpha\), Ly\(\alpha\) and FUV
continuum as well as HST COS spectroscopy of the same galaxies.
Additionally, the study encompasses observations in H {\sc i} from the Green
Bank Telescope (GBT) and the Karl I. Jansky Very Large Array (VLA);
multi-wavelength optical observations from the Nordic Optical Telescope
(NOT), Calar Alto PMAS IFU observations, HST-COS FUV spectroscopy
presented here and treated in further detail by Orlitova et al. (in
prep.), as well as archival data from SDSS, GALEX, and more.

In this paper, we present a detailed analysis of kinematics and geometry of the
neutral medium along the line of sight towards the brightest star-forming
regions of the LARS galaxies, inferred from absorption features in the UV
continuum as observed in spectroscopy performed with the HST COS under programs
11522, 11727, 12027, and 12583. The aim of this analysis is to investigate how
these properties influence the transport and escape of Ly\(\alpha\) in these
galaxies. We are, for the first time, able to compare our spectroscopic
findings to global properties of the sample galaxies, like e.g.~global
Ly\(\alpha\) escape fraction, extracted and computed from HST imaging as
presented in \citet{LARSII}. In this paper, ``global'' properties of a galaxy
means properties of the full galaxy as obtained from either HST imaging or 21 cm
radio interferometry or single-dish observations, presented in \citet{LARSII}
and \citet{LARSIII}, whichever is appropriate from context. We measure a
selection of metal lines in the low-ionized state (LIS) medium. LIS metal
absorption lines are well established as tracers of neutral gas \citep[see
e.g.][]{Pettini2002, Erb2012, Sandberg2013}, and have been employed in studies
of relations between neutral ISM kinematics/geometry and Ly\(\alpha\) radiative
transfer \citep{Kunth98, Shapley2003, Jones2013} and Lyman Continuum escape
\citep{Heckman2011, Jones2013, JaskotOey}. In particular, we focus on the four
Si {\sc ii} lines at \(\lambda \lambda\) 1190, 1193, 1260 and 1304 to
investigate in- or outflowing winds, line strengths, and utilize the fact that
these lines all arise form the same ground state to attempt to resolve the
ambiguity of whether residual flux at line center stems from low column density
or low covering factor.  Furthermore, we can compare our findings to global H
{\sc i} properties inferred from 21 cm GBT single dish and VLA interferometry
observations as presented in \citet{LARSIII} and in a broader scope the findings
of other, related works in the field of Ly\(\alpha\) radiative transfer.

Furthermore, we stack and measure the Si {\sc iv} transitions at
\(\lambda \lambda\) 1122, 1393 and 1402 to trace the hot, high-ionized
medium and compare its properties to those of the neutral medium.

The structure of this paper is as follows: In Section~\ref{sec:obs}, we
briefly present the sample, describe the observations involved in this
analysis and the data reduction and preparations. In
Section~\ref{sec:measurements}, we present data calibration and measurements,
and give a short description of the methods involved. In
Section~\ref{sec:results}, we present the results of these measurements and
the subsequent analysis and comparisons with other works involving the
sample. In this section, we also discuss how to interpret our findings.
General discussion of our findings, comparisons to previous research in the
field and possible caveats is done in Section~\ref{sec:discuss}. A summary
and conclusion is given in Section~\ref{sec:conclusion}, and a discussion of
individual targets is given in Appendix \ref{sec:targets}.

\section{Observations and data}\label{sec:obs}

\subsection{Sample}\label{sample}

A full description and motivation of the sample is found in
\citet{LARSI}. Here, we provide a short summary.

Since the basic idea of the Lyman Alpha Reference Sample is to
investigate the fate of Ly\(\alpha\) photons, which are mainly produced
in the hot gas surrounding star/forming regions, the sample should
consist of galaxies of strong star formation. To achieve this, a
selection was made of UV-bright galaxies from the SDSS and GALEX
catalogs. The galaxies were chosen to span UV-luminosities similar to
those of Ly\(\alpha\) emitters and Lyman Break Galaxies (LBGs) at
redshifts \(\sim 3\). To ensure that the sample consists of actively
star-forming galaxies, LARS galaxies are selected to have a H\(\alpha\)
equivalent width of W(H\(\alpha\))\(>100 \AA\), a domain in which the
galaxy luminosity is dominated by the young population of OB stars and a
high production of Ly\(\alpha\). Galaxies with FWHM(H\(\alpha\))
\(> 300\) km s\(^{-1}\), indicating an AGN component, were rejected. The
galaxies have redshifts within a range of \(0.02 < z < 0.2\); the lower
side keeps Ly\(\alpha\) well clear of the geocoronal Ly\(\alpha\) line,
while the higher side keeps H\(\alpha\) within the imaging bandpass of
the ACS camera on the HST.

LARS overlaps with other COS campaigns. LARS 1, 6, 7 and 10 are observed
for and included in the sample of \citet{Wofford2013} as KISSR 1578,
2019, 242 and 178, respectively. LARS 12, 13 and 14 are observed by
\citet{Heckman2011}. The COS data of LARS 7 have also been published in
\citet{France2010} as KISSR 242.

LARS galaxies have been observed in several wavelength ranges and with
various techniques: In UV continuum-, Ly\(\alpha\)- and H\(\alpha\)
imaging with the HST \citep{LARS0, LARSI, LARSII}, in 21 cm H {\sc i} emission
with Green Bank Telescope single dish and Karl I. Jansky VLA radio
interferometric observations \citep{LARSIII}, and Calar Alto PMAS IFU
observations. An overview of key numbers describing the sample can be
found in Table~\ref{tab:sample}, including high precision redshifts
derived from SDSS DR7 archival data.

\begin{deluxetable*}{cccrc}
\tablecolumns{5}
\tablewidth{0pt}
%\tabletypesize{\scriptsize}
\tablecaption{Basic properties of the LARS galaxies.\label{tab:sample}}
\tablehead{\colhead{ID} & \colhead{Common name} & \colhead{$z$ (measured)} &
           \colhead{RA} & \colhead{Dec} \\
	   \colhead{} & \colhead{} &\colhead{} & \colhead{[h:m:s]} 
	   &\colhead{[d:m:s]} \\
           \colhead{(1)} & \colhead{(2)} & \colhead{(3)} & \colhead{(4)} & 
	   \colhead{(5)}
	   }
\startdata
LARS 01 & Mrk 259        & \(0.027967 \pm 2.6\times 10^{-5}\) & 13:28:44.161 & +43:55:51.31 \\
LARS 02 & \ldots{}       & \(0.029836 \pm 1.8\times 10^{-5}\) & 09:07:03.150 & +53:27:13.28 \\
LARS 03 & Arp 238        & \(0.030733 \pm 6.5\times 10^{-5}\) & 13:15:32.983 & +62:07:43.89 \\
LARS 04 & \ldots{}       & \(0.032530 \pm 1.8\times 10^{-5}\) & 13:07:26.330 & +54:27:09.97 \\
LARS 05 & Mrk 1486       & \(0.033856 \pm 5.6\times 10^{-5}\) & 13:59:48.958 & +57:26:38.11 \\
LARS 06 & KISSR 2019     & \(0.034157 \pm 3.1\times 10^{-5}\) & 15:45:43.070 & +44:16:09.82 \\
LARS 07 & IRAS 1313+2938 & \(0.037808 \pm 2.5\times 10^{-5}\) & 13:16:02.420 & +29:23:15.57 \\
LARS 08 & \ldots{}       & \(0.038185 \pm 2.3\times 10^{-5}\) & 12:50:12.483 & +07:35:04.10 \\
LARS 09 & IRAS 0820+2816 & \(0.047194 \pm 1.2\times 10^{-5}\) & 8:23:53.722  & +28:06:38.59 \\
LARS 10 & Mrk 0061       & \(0.057381 \pm 2.3\times 10^{-5}\) & 13:01:40.301 & +29:23:10.09 \\
LARS 11 & \ldots{}       & \(0.084416 \pm 5.5\times 10^{-5}\) & 14:03:46.251 & +06:28:28.05 \\
LARS 12 & SBS 0934+547   & \(0.102075 \pm 4.7\times 10^{-5}\) & 09:38:11.650 & +54:28:41.94 \\
LARS 13 & IRAS 0147+1254 & \(0.146703 \pm 4.7\times 10^{-5}\) & 01:50:27.210 & +13:09:14.96 \\
LARS 14 & \ldots{}       & \(0.180691 \pm 5.6\times 10^{-5}\) & 09:25:58.924 & +44:27:53.29 
\enddata
\tablecomments{For other quantities, see \citet{LARS0}, \citet{LARSII},
\citet{LARSI}.}
\end{deluxetable*}

\subsection{Observations and data reduction}\label{sec:observations}

As described in Section~\ref{sec:obs}, the LARS sample overlaps with two
other samples for which COS spectroscopy has been obtained in the
appropriate settings. Here we rely upon the archival data, and GO
program ID numbers are listed in column 2 of Table~\ref{tab:observations}.
Specifically, LARS targets 1, 6, 7, and 10 were observed during COS
guaranteed time observations \citep[ID: 11522,12027, PI: Green,
see][]{Wofford2013}, and LARS 12, 13, and 14 were observed under GO
11727 \citep{Heckman2011}.

Both sets of archival observations were obtained using the COS
acquisition modes \texttt{ACS/IMG} or \texttt{ACS/SEARCH}, which
successfully acquired and centred upon the brightest NUV point that fell
within the Primary Science Aperture (PSA). For the FUV spectroscopy,
somewhat different strategies have been adopted by the individual
observers. GO 11522,12027 used one orbit per target, obtaining spectra
in the G130M grating. These observers used two different central
wavelength settings (\texttt{CENWAVE}=1291 and 1318 throughout; column 3
of Table~\ref{tab:observations}) to perform a spectroscopic dither that
spanned the gap between the two segments of the detector. GO 12727
observations instead used 2 or 4 orbits per target and, because of the
larger redshift of these galaxies, obtained spectra in both G130M and
G160M gratings. In each grating two exposures were obtained using
different \texttt{FP-POS} positions of the grating to perform a small
spectral dither.

Our own observations (GO 12583; LARS 02, 03, 04, 05, 08, 09, 11) also used one
orbit per target. We first used the ACS/SBC/F140LP and WFC3/UVIS/F336W, obtained
as part of the imaging campaign \citep{LARSI} to identify the regions of peak UV
surface brightness. Knowing precisely the target coordinates for the COS
observations we acquired the targets with \texttt{ACQ/IMG} with the NUV imager.
Again targets were perfectly centred to within a couple of NUV pixels. For five
of the seven targets we noted that multiple UV sources would fall within the
PSA, and may degrade spectral resolution, so in these cases we imposed orient
constraints (\(\pm 20\deg\)) in order to align multiple sources with the
cross-dispersion axis. We performed spectral observations with the G130M
grating; since the targets vary in redshift between \(z=0.028\) and 0.084, we
chose the \texttt{CENWAVE} settings (again see Table~\ref{tab:observations}) to
include Ly\(\alpha\) and the largest possible number of low-ionization
absorption lines, tuning the spectral gap between the detector segments to
regions without features of interest.  For each target we performed spectral
dithers using three \texttt{FP-POS} settings, except for LARS 3 which was
observed in the continuous viewing zone (CVZ) and used all four \texttt{FP-POS}
positions.

\begin{figure*}[htbp]
\centering
\includegraphics{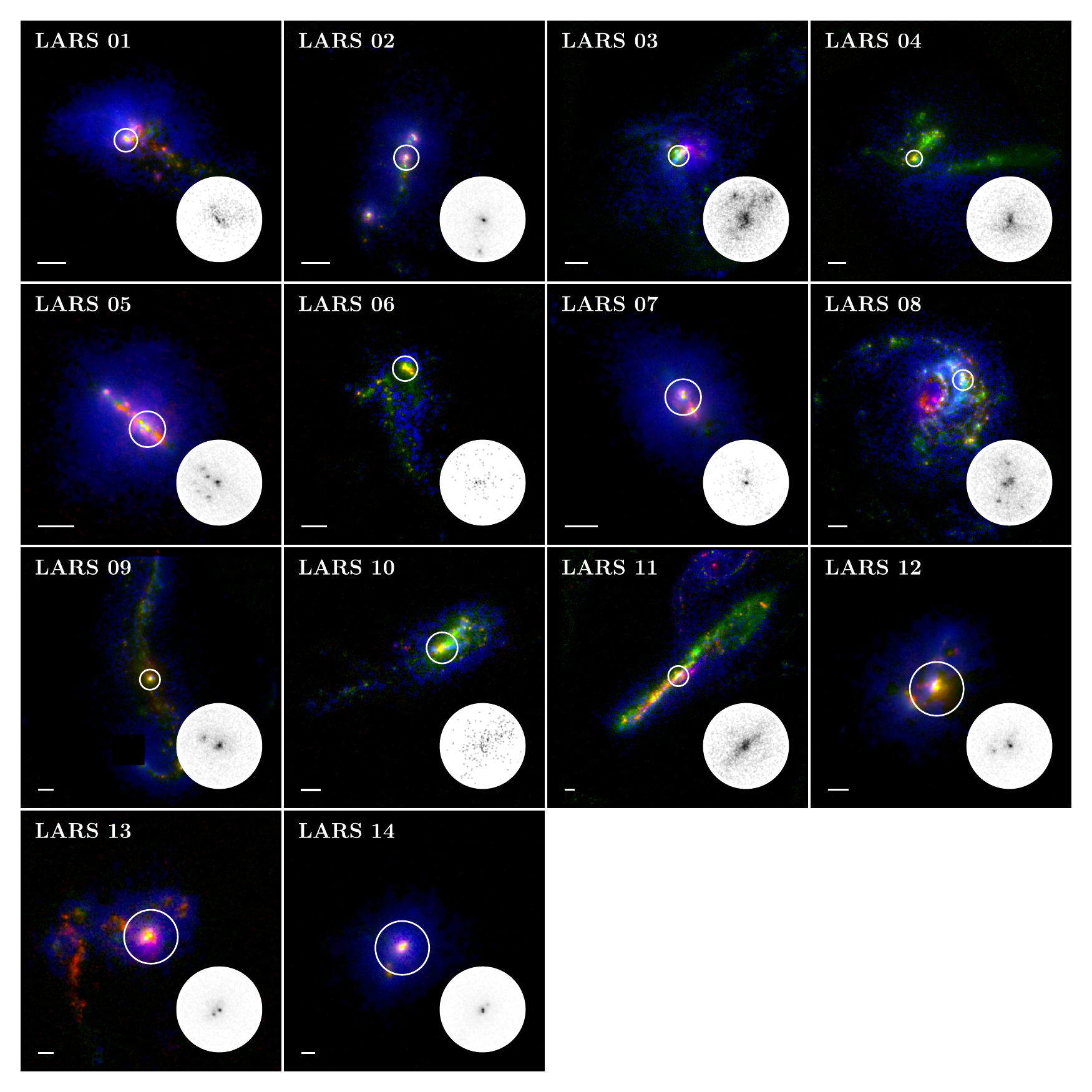}
\caption{The COS aperture (white circles) overlaid on composite RGB images of
the LARS galaxies in UV-continuum (green), H\(\alpha\) (red) and Ly\(\alpha\)
(blue), as seen in \citet{LARSII}. The aperture is pointed at brightest
star-forming knot in each galaxy as seen in the UV-continuum. Inset in the lower right corner of each panel is shown, in reverse grayscale, the galaxy as seen through the COS aperture by the acquisition imaging camera.  White scale bars in the lower left of each panel shows 1 kpc. All images are aligned so up is north, left is east. Note that the Ly\(\alpha\) emission only traces the H {\sc i} halo of the galaxy to the extent that it escapes from the inner parts of the galaxy; absence of emission does not imply absence of H {\sc i} gas. The COS aperture only captures a fraction of global Ly\(\alpha\) emission.
\label{fig:apertures}}
\end{figure*}

For illustration we show the position of the PSA (radius 1.25\arcsec)
on each target in Figure~\ref{fig:apertures}. In each case this found
the brightest star cluster in the FUV. Images are those of
\citet{LARSII}, encoding continuum-subtracted H\(\alpha\) in red, FUV
continuum in green, and continuum-subtracted Ly\(\alpha\) in blue. The
PSA is shown as a circle in each image. The lower right inset of each
panel shows the corresponding COS/NUV acquisition frame, masked to match
the aperture.

\begin{deluxetable*}{ccccrc}
\tablecolumns{6}
\tablewidth{0pt}
%\tabletypesize{\scriptsize}
\tablecaption{Key information about observations.\label{tab:observations}}
\tablehead{\colhead{ID} & \colhead{Prog. ID}& \colhead{Grating} &
           \colhead{Cenwave} &
           \colhead{Exp.\ time} & \colhead{$R_{\mathrm{eff}}$}  \\
	   \colhead{} & \colhead{} &\colhead{} & \colhead{[\AA]} & 
	   \colhead{[s]} &\colhead{[resel]}\\
           \colhead{(1)} & \colhead{(2)} & \colhead{(3)} & \colhead{(4)} & 
	   \colhead{(5)} & \colhead{(6)}
	   }
\startdata
LARS 01 & 11522 & G130M       &1291,1318                   & 1153.184 & 2.11  \\
LARS 02 & 12583 & G130M       &1327                        & 763.168  & 1.64  \\
LARS 03 & 12583 & G130M       &1327                        & 1005.216 & 2.57  \\
LARS 04 & 12583 & G130M       &1327                        & 785.184  & 2.57  \\
LARS 05 & 12583 & G130M       &1327                        & 722.176  & 1.40  \\
LARS 06 & 11522 & G130M       &1291,1318                   & 1142.208 & 2.57  \\
LARS 07 & 11522 & G130M       &1291,1318                   & 1111.168 & 2.11  \\
LARS 08 & 12583 & G130M       &1327                        & 701.184  & 3.04  \\
LARS 09 & 12583 & G130M       &1309                        & 618.176  & 1.64  \\
LARS 10 & 12027 & G130M       &1291,1318                   & 1105.216 & 3.27  \\
LARS 11 & 12583 & G130M       &1300                        & 698.208  & 3.04  \\
LARS 12 & 11727 & G130M,G160M &1291,1600 \tablenotemark{a} & 1140.192 & 1.40  \\
LARS 13 & 11727 & G130M,G160M &1327,1623 \tablenotemark{a} & 1800.192 & 1.17  \\
LARS 14 & 11727 & G130M,G160M &1300,1611 \tablenotemark{a} & 2550.208 & 0.70  
\enddata
\tablenotetext{a}{Settings for gratings G130M and G160M, respectively.}
\tablecomments{Col.~6 lists effective resolutions in
units of COS resolution elements as described in Sec.~\ref{sec:resolution}.}
\end{deluxetable*}

Data reduction was performed with \texttt{CALCOS} version 2.15.6
(2011-11-03) pipeline tools. Because our absorption features do not fall
close to geocoronal emission lines we make no attempt to correct for O {\sc i}
lines near 1302 Å as done in \citet{James2014}, instead collecting
continuum photons obtained at all Earth-limb angles. Finally data were
resampled onto a uniform wavelength grid using the \(\Delta \lambda\)
corresponding to the coarsest value, and stacked rejecting any pixel for
which the data quality flag was \(>0\). We check for systematic offsets
in the wavelength calibration by measuring the centroids of the
geocoronal emission lines, using the wings of Ly\(\alpha\) and the O {\sc i}
\(\lambda 1302\) line. For this we produced additional combinations of
the extracted spectra, which permitted the inclusion of detector pixels
with low pulse height amplitude and gain sag, so as to fully include the
geocoronal emission. In no case did we measure the wavelength of
geocoronal emission features to be offset from the systemic values by
more than the resolution; even accounting for switching between mirrors
B and A in target acquisition (LARS 05 and 09 only) we observe
no reason to doubt the wavelength calibration within about 20 km
s\(^{-1}\).

\begin{figure*}[htbp]
\centering
\includegraphics{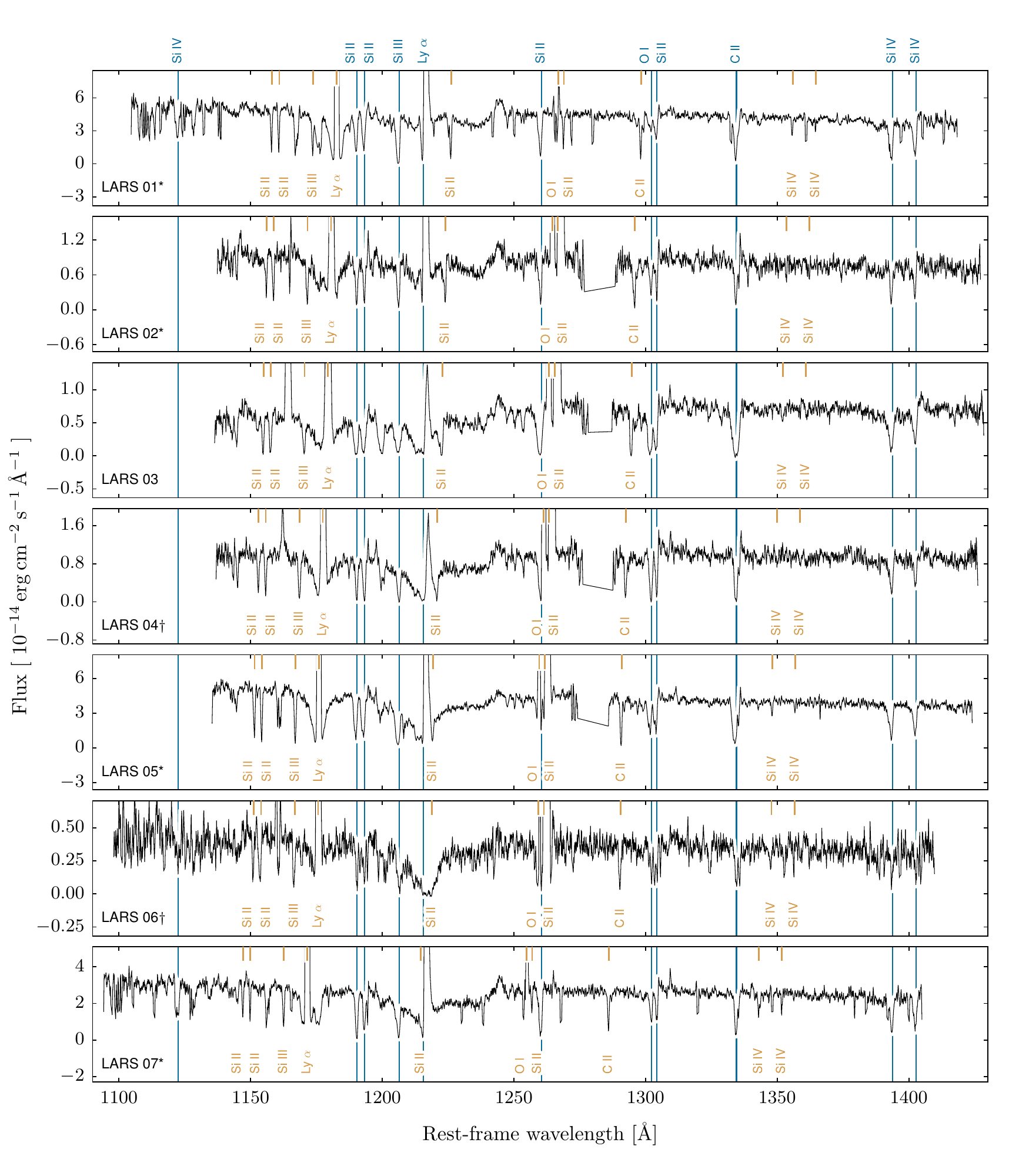}
\caption{Full COS spectra of LARS 1 through 7 in units of rest-frame
wavelength. Features included in our analysis are shown with blue, while
Milky Way and geocoronal features are marked in orange. Objects marked
with $*$ denote that the object has a global Lyman-$\alpha$ escape fraction
$\gtrsim 10\%$ as found from HST photomotry by \citet{LARSII}. Objects marked
with a $\dagger$ are global Ly$\alpha$ absorbers.
\label{fig:fullspec1-7}}
\end{figure*}

\begin{figure*}[htbp]
\centering
\includegraphics{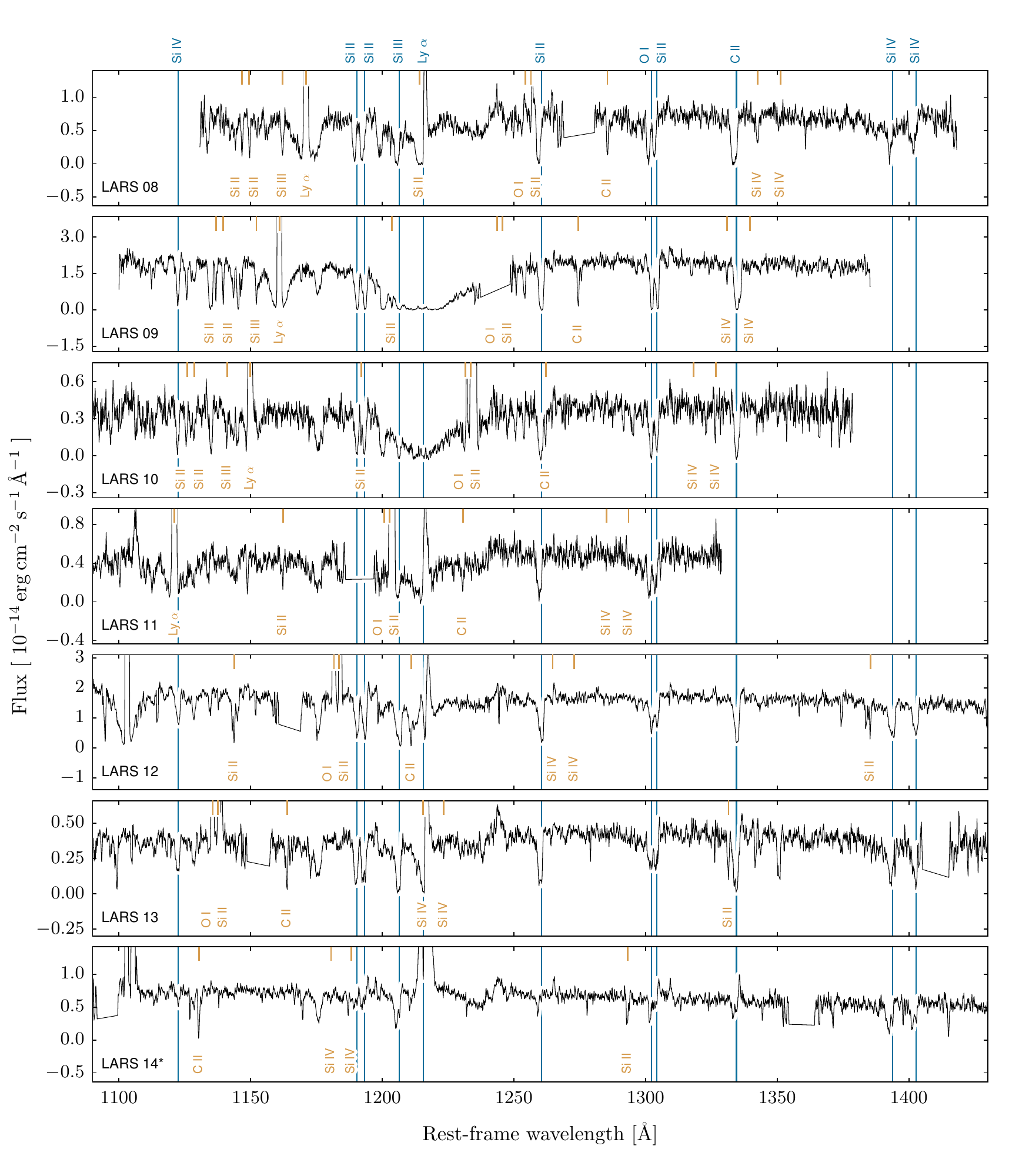}
\caption{Same as Figure~\ref{fig:fullspec1-7}, but for LARS 8 through 14.
\label{fig:fullspec8-14}}
\end{figure*}

Figures~\ref{fig:fullspec1-7} and~\ref{fig:fullspec8-14} show the full
spectra of all 14 galaxies in their rest frame. The transitions analyzed
in this study are marked in blue; geocoronal and MW features are shown
in orange. The spectra cover varying wavelength ranges; some spectra
have been cut off where they did not contain any information included in
this analysis. Galaxies with global Ly\(\alpha\) escape fraction
\(> 10\%\) in \citep{LARSII} are marked with a \(\ast\), global
Ly\(\alpha\) absorbers are marked with a \(\dagger\).

\section{Measurements}\label{sec:measurements}

\subsection{Systemic redshift}\label{sec:systemic}

We base the systemic redshift determination on nebular emission lines in the
hot regions surrounding the starbursts. We assume these lines originate from
the same recombination processes as Ly\(\alpha\) and therefore can be expected
to trace the intrinsic Ly\(\alpha\) well.  While no strong nebular emission
lines other than Ly\(\alpha\) are present in our COS spectra, such lines can be
found in the SDSS archival spectra. Systemic redshifts were determined as
described in \citet{LARSI} by aid of the lines H\(\alpha, \beta, \gamma
\textrm{\ and\ } \delta\); {[}O {\sc i}{]} \(\lambda 6300\), {[}O {\sc iii}{]}
\(\lambda \lambda 4959,5007\), {[}N {\sc ii}{]} \(\lambda \lambda 6548, 6583\)
and {[}S {\sc ii}{]} \(\lambda \lambda 6716, 6731\).

The COS pointings of this sample are very close to those of SDSS; we do not
report the numbers here but for the subsample included in \citet{Wofford2013},
the offsets have been measured to 0.10, 0.08, 0.38 and $0.62\arcsec$ for LARS 1,
6, 7 and 10, respectively. The aperture sizes are similar as well,
\(2.5\arcsec\) for
COS vs. \(3.0\arcsec\) for SDSS, so the systemic redshift derived from SDSS
spectroscopy is believed to describe the same regions from which the
Ly\(\alpha\) photons seen in the COS spectra originate. These redshifts were
found by fitting each of the above mentioned lines in the SDSS data to a single
Gaussian, inferring a redshift from each and since take a S/N-weighted average
of these redshifts. SDSS-derived redshifts are listed in
Table~\ref{tab:sample}; other properties of the LARS galaxies measured and
derived from SDSS spectroscopy can be found in \citet{LARSI}. The computed
redshifts all agree with those listed in the SDSS database to within \(\sim
10\, \rm km\, s^{-1}\).

In the case of LARS 01, 06, 07, and 10, (KISSR 1578, 2019, 242 and 178,
respectively), the systemic redshifts were found to agree with those
reached by \citet{Wofford2013} to within the precision reported there.

\subsection{Effective resolution}\label{sec:resolution}

The COS has a resolution of \(R = 20,000\) for a perfect point source, which
corresponds to 6 detector pixels in the dispersion direction. We rebinned the
data by a factor of 6 to reflect this, improving S/N without losing information.

The targets of this study are not point sources, so the effective
resolution is somewhat lower than the \textbf{instrument resolution}, the
amount depends on the extent and morphology of each target. We estimated the
effective resolution for each target based on the reduced and cleaned UV
continuum frames prepared by \citet{LARSII} in the following way:

We created a circular mask on the FUV frame corresponding to the COS
aperture, rotated the part of the frame inside the mask to align with
the COS acquisition image. Vignetting was taken into account by
multiplying the flux in each aperture pixel with an interpolation of the
aperture throughput grid in Figure~7 of \citet{CosImaging}, and the flux
then summed along the cross-dispersion direction. The half-light width
of the resulting profile was found by linearly interpolating on a grid
ten times finer than the actual pixel scale. The resulting width in
pixel units was converted to arcsec, which again can be translated to
COS resolution elements by the conversion factor \(0.171\) arcsec/resel
reported in the COS Instrument Handbook \citep{CosHandbook}. The
resulting widths are listed in Col.~6 of Table~\ref{tab:observations}

\subsection{Metal absorption line profiles}\label{sec:lipro}

Key properties of the absorption lines selected for analysis in this work are
tabularized in Table~\ref{tab:lines}. Four Si {\sc ii} absorption lines, one O
{\sc i} and one C II feature are together assumed to trace the neutral medium.
The Si {\sc ii} lines all arise from the ground state in Si II, allowing us to
utilize their difference in line strength to evaluate Si \textsc{ii}
column density and neutral gas covering factor.

Three Si {\sc iv} lines are presented and briefly analyzed; they are
believed to trace the interface between the hot, ionized medium in the regions
around the starburst, and the surrounding neutral medium  \citep{Grimes2009,
Strickland2009}.  These lines may also contain a stellar component.

Finally, fluorescent emission lines connected to the fine structure splitting in
the ground state of Si {\sc ii} are included in the table, as they hold some
significance to the discussion of our results.  These lines arise when
electrons in the excited state of a transition with significant fine-structure
splitting of the ground state decay to the upper of these levels, which then
quickly after decays into the lower fine structure level by emission of a
long-wavelength photon. Since the upper level of the ground state is generally
not populated, they are seen in emission only as a peak somewhat redwards of the
main, resonant absorption feature - in case of the silicon lines treated here,
the separation is $\sim 4$ \AA. These lines are marked with a $^*$ in
Table~\ref{tab:lines}.

\begin{deluxetable}{rcccr@{}}
	\tablecaption{Properties of transitions analyzed.\label{tab:lines}}
\label{tab:trans}
\tablehead{
\colhead{Transition} & \colhead{\(\lambda_0\) [Å]} & 
\colhead{\(\mathrm{f}_{ik}\)} & \colhead{\(E_{\mathrm{ion}} [\rm keV]\)} &
\colhead{\(f_{\lambda} \)[Å]}\\
\colhead{(1)} & \colhead{(2)} & \colhead{(3)} & \colhead{(4)} & \colhead{(5)}
}
\startdata
Si \textsc{iv } 1122 & 1122.48490 & 0.8190 & 45.142 & 919.3151\tabularnewline
Si \textsc{ii } 1190 & 1190.41580 & 0.2820 & 16.346 & 335.6973\tabularnewline
Si \textsc{ii } 1193 & 1193.28970 & 0.5850 & 16.346 & 698.0745\tabularnewline
Si \textsc{ii*} 1194 & 1194.500   & 0.7370 & 16.346 & 880.3465\tabularnewline
Si \textsc{ii*} 1197 & 1197.390   & 0.1500 & 16.346 & 179.6085\tabularnewline
Si \textsc{ii } 1260 & 1260.42210 & 1.2300 & 16.346 & 1550.3192\tabularnewline
O  \textsc{i  } 1302 & 1302.16848 & 0.0480 & 13.618 & 62.5041\tabularnewline
Si \textsc{ii } 1304 & 1304.37020 & 0.0925 & 16.346 & 120.6542\tabularnewline
C  \textsc{ii } 1334 & 1334.53230 & 0.1280 & 24.383 & 170.8201\tabularnewline
Si \textsc{iv } 1393 & 1393.75460 & 0.5080 & 45.142 & 708.0273\tabularnewline
Si \textsc{iv } 1402 & 1402.76970 & 0.2520 & 45.142 & 353.4980
\enddata
\tablecomments{Column 2 holds the restframe wavelength of each
transition. Column 3 holds the dimensionless oscillator strength $f_{ik}$
of the transition. Column 4 holds the species' ionisation energy in keV, and
column 5 holds the product of oscillator strength and rest wavelength.}
\end{deluxetable}

After reduction and resampling, the spectra were continuum-normalized
around the absorption features of interest. This normalizing was
performed by hand, as the continuum is generally not sufficiently
well-behaved to be modeled. For each absorption feature, the surrounding
wavelength ranges showing values which were deemed to make up the
continuum were selected by hand, after which these values were fitted to
a simple linear function, by which the data were then normalized.
Several of the spectra show blending of the lines included in our
analysis with MW and geocoronal features. Wavelength ranges affected by
this were masked out and not included in the analysis.

The normalized lines and immediate surroundings were then interpolated
linearly and evaluated on the coarsest grid in velocity space, i.e.~the
velocity step size of the line with lowest rest wavelength of each
galaxy.

\begin{figure*}[htbp]
\centering
\includegraphics{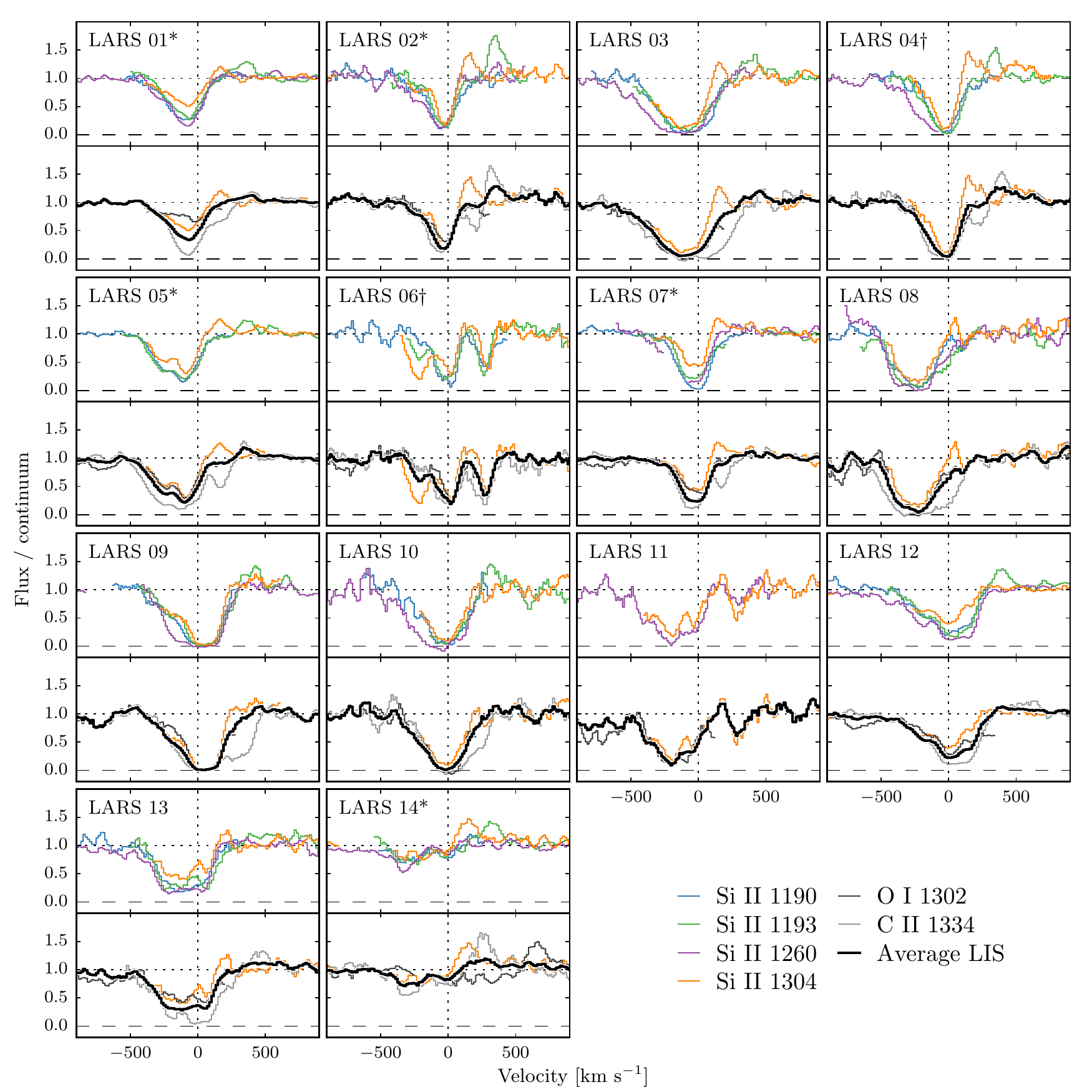}
\caption{The absorption line profiles in velocity space of the Si {\sc ii}
transitions at \(\lambda \lambda 1190, 1193, 1260, 1304\) for each of
the 14 LARS galaxies. The zero-point is derived from the nebular
emission lines of the SDSS spectra of \citet{LARSI} The lines are
smoothed by a box kernel for illustrative purposes. Regions masked out
because of contaminating features are not shown. \label{fig:liproplot}}
\end{figure*}

Figure~\ref{fig:liproplot} shows in the upper panel for each galaxy the
normalized and resampled absorption profiles in velocity space of the Si
II transitions at \(\lambda \lambda\) 1190, 1193, 1260 and 1334. Masked
out wavelength ranges are not shown. The zero point value is the
systemic velocity derived from SDSS spectroscopy of the nebular lines of
the H {\sc ii} regions surrounding galaxies' star-forming regions. The profiles
show a broad range of velocity shifts relative to systemic, as well as a
variety of widths, depths and shapes. The lower panels of Figure~\ref{fig:liproplot} show, for reference, the profile of Si {\sc ii}
\(\lambda\) 1304 (the only one of the transitions present in all 14 LARS
galaxies) along with the profiles of O {\sc i} \(\lambda 1302\) and C II
\(\lambda 1334\). Also, in thick black, is shown the average of these
and the Si {\sc ii} lines or the subset hereof available in the given galaxy.
Data shown in Figure~\ref{fig:liproplot} have been smoothed by a flat
kernel 5 bins wide for clarity.

Despite their different oscillator strengths, there is a tendency for the lines
to have almost equal depth and shape. This is usually an indicator that the
profiles are the result of absorption in an optically thick medium with a
less-than-unity covering fraction; For a fully covering but part
transparent gas screen, transitions of different strengths would show a
corresponding different in depth of their absorption features (see
Sect.~\ref{sec:covfrac}).

\begin{figure*}[htbp]
\centering
\includegraphics{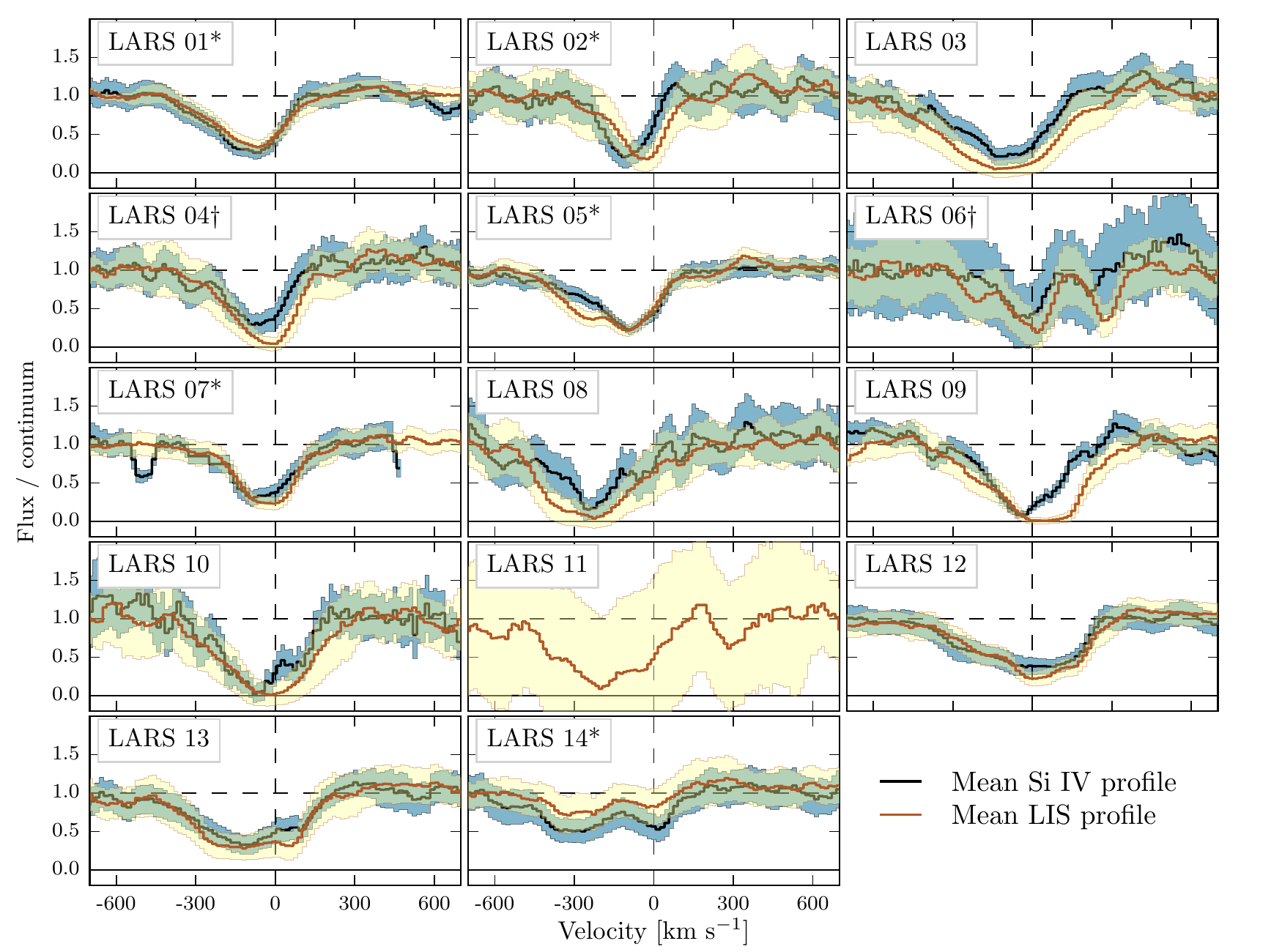}
\caption{Averaged profiles of the neutral and ionized phases. Shaded
areas signify standard errors. Both profiles and errors are smoothed for
illustrative purposes.\label{fig:phases}}
\end{figure*}

Like the LIS lines, the Si {\sc iv} transitions were resampled in velocity
space as described above and averaged for comparison with the LIS lines.
These are shown in Figure~\ref{fig:phases} with their propagated
standard errors as a surrounding shaded region.

\subsection{Velocity binned neutral ISM covering}\label{sec:covfrac}

The presence of residual flux in a given velocity interval within an
absorption feature can indicate either that the gas at these velocities
is not completely opaque, or that this gas is opaque but only partly
covering the background light source; or a combination hereof. In the
presence of multiple lines arising from the same ground state, this can
be disentangled by utilizing the differences in oscillator strengths
\(f\lambda\) of said lines in the \emph{apparent optical depth method}
described by \citet{Savage1991}, used by e.g. \citet{Pettini2002} and
\citet{Quider2009}. The basic idea behind this method is that the
different lines, in the case of an optically thin medium, will show
different absorption depths depending on their oscillator strengths,
while an optically thick system with partial covering will yield
identical profiles for the different transitions. An extension of the
method is employed in this work, which allows us to compute the column
density and covering fraction of these lines as a function of velocity
offset from the systemic zero-point \citep{Jones2013}. For want of any
suitable transitions of H {\sc i} to directly measure the covering fraction of
the neutral ISM, we utilize the presence of four Si {\sc ii} lines within the
wavelength range of our spectra. The method is based on the following.

For a given velocity range, the column density \(N\) and the covering
fraction \(f_C\) can be calculated from the ratio of residual intensity to
continuum level by the following relation:

\begin{equation}
\frac{I}{I_0} = 1 - f_C (1 - e^{- \tau }).
\end{equation}

The optical depth \(\tau\) is related to the column density and
oscillator strength as:

\begin{equation}
\tau = f \lambda \frac{\pi e^2}{m_e c}N = f \lambda \frac{N}{3.768 \times
10^{14}}, \label{eq:tau} 
\end{equation}

\noindent where \(f\) is the oscillator strength, \(\lambda\) the wavelength 
in \(\AA\), and \(N\) the column density (in the given velocity range) in
cm\(^{-2}\). In the presence of multiple absorption lines with different
values of \(f \lambda\), arising from the same energy level in the same
species, we can fit the data for \(N\) and \(f_C\) for each velocity bin.
Our spectra cover the wavelengths of four lines arising from the ground
state of Si {\sc ii} with the wavelengths \(\lambda \lambda 1190, 1193,
1260\) and 1304.

Practically, this computation was performed by formulating \(I/I_0\) as
a function of \(f \lambda\) and finding the values of \(N\) and
$f_C$ that minimize the squared residuals of
subtracting the measured \(I/I_0\) from the theoretical ones. The
procedure for the minimization was to create a simple grid of
\((N, f_C)\) value pairs limited by the physically sensible, find the
squared residuals in each point and then finding the combination that
minimized this function. Confidence levels for the two parameters were
found by projecting the \((\chi_{\rm min}^2+1)\) contour (see below)
onto the \(N\) and \(f_C\) axes.

It is important to note at this point that the values of \(f_C(v)\)
found in this way is \emph{not} what one would usually call the
``covering fraction'' of the neutral gas. The \(f_C\) are the covering
fractions of individual slices in velocity space of the gas, while
covering fraction in the usual sense of this term, from here on denoted
\(F_C\) with a capital \(F\), is the covering factor of all neutral gas
regardless of velocity. Since the individual velocity-space slices
generally do not occupy the same projected physical space, it is
entirely possible to have \(f_C < 1\) everywhere while still having a
total \(F_C \simeq 1\); the found values of \(f_C\) provide only a lower
limit for \(F_C\).

\subsubsection{Fit quality indication}\label{sec:fitqual}

Due to contamination by Milky Way lines and geocoronal lines, lines
falling outside of detectors at certain redshifts etc., not all
transitions have usable data in all velocity bins for all galaxies. The
minimum number of contributing lines required for our fitting procedure
is 2. In this case, with two data points and two free parameters, there
are zero degrees of freedom and thus the usual reduced \(\chi^2\) is not
defined. Instead, we report a pseudo-reduced \(\chi^2\), defined as
\(\chi^2 / ([DOF]+1)\). This goodness-of-fit indicator is defined
everywhere, has the same minimum as the normal \(\chi^2\), and gives
similar but not identical error estimates.

\subsubsection{Averaged absorption
profiles}\label{averaged-absorption-profiles}

Additionally, also in analogy with \citet{Jones2013}, we have utilized a
simple, complementary method to estimate the covering fractions by
investigating saturated transitions. In the case where \(\tau \gg 1\),
equation~\ref{eq:tau} reduces to:

\begin{equation}
I/I_0 = 1-f_C
\end{equation}

In this case, we would expect all reasonably strong lines to show the
same absorption profile, and said profile will be a good proxy for the
covering fraction. The Si {\sc ii} absorption profiles in Figure~\ref{fig:liproplot} suggest that the four Si {\sc ii} lines are indeed
saturated, as does similar plots of the lines of O {\sc i} \(\lambda 1302\)
and C II \(\lambda 1334\).

To minimize random variations, an inverse variance weighted average of
the low-ionized absorption profiles is used for the further analysis; as
usual, contaminated regions are masked out and not included in the final
profile.

\begin{figure}[htbp] \centering \includegraphics{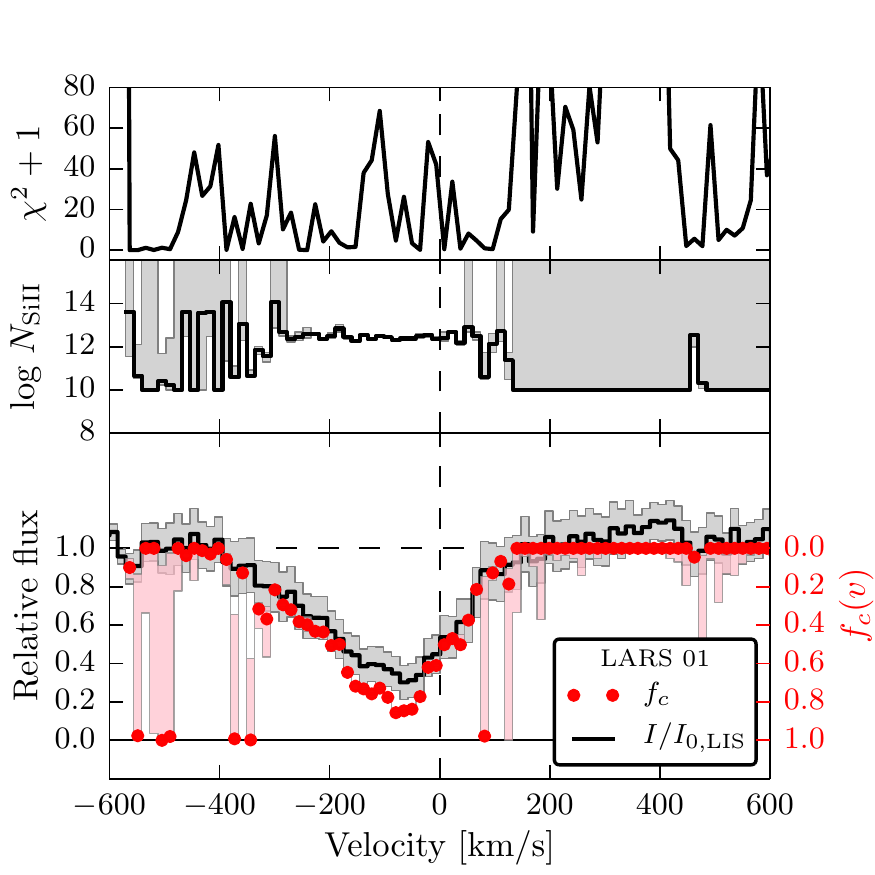}
\caption{\textbf{Upper panel:} Pseudo reduced \(\chi^2\) as described in
Section~\ref{sec:fitqual}. \textbf{Middle panel:} \(\log N_{\rm Si
\textsc{ii}}\) as computed by the method described in~\ref{sec:covfrac} in
black, with \((\chi^2_{\min} +1)\) confidence levels in shaded gray, cut off at
13 and 20, marking the limits of the grid described in
Section~\ref{sec:covfrac}. \textbf{Lower panel:} \emph{Black steps:} Average
low-ionisation absorption profile with standard errors in shaded gray.
\emph{Red dots:} Computed covering fraction for each bin, with pink shading
indicating the \(\chi^2_{\min}+1\) confidence levels.  \label{fig:cofral1}}
\end{figure}

Figure~\ref{fig:cofral1} shows the averaged absorption profile of LARS 1
with the computed covering fraction overlaid, along with the computed
column densities on logarithmic scale and the pseudo-reduced \(\chi^2\)
described in Section~\ref{sec:covfrac}. It is apparent that the averaged
line profile does indeed follow the computed covering fractions quite
closely and thus is indeed a good proxy for the actual covering
fraction. The other galaxies of the sample show similar behavior, as can
be seen in Figures~\ref{fig:covfracs} through~\ref{fig:coverfracs3}.

The same figures also show that these averaged profiles have
considerably less scatter than the computed values of \(f_C(v)\). Thus,
after having shown that they do indeed follow the \(f_C(v)\) profiles
well, we can benefit from adapting the averaged profiles as proxies for
these in our further computations, although with some words of caution
for a few of the galaxies (see Appendix~\ref{sec:targets}).

\subsection{Velocities}\label{sec:velocities}

When describing the kinematics of the neutral gas, a good set of characteristic
velocities is needed. The approach usually taken for simple line profiles is to
approximate them to a simple or possibly skewed Gaussian or Voigt profile and
describe these by their centroid and FWHM or, for the Voigt profile, the
broadening parameter \(b\).  However, our profiles generally are too complex and
asymmetric to be well approximated by a single profile. Instead, we characterize
the profiles by a set of velocities based on their integrated area as follows:

\begin{description}
\itemsep1pt\parskip0pt\parsep0pt
\item[Integrated central velocity (\(v_{\rm int}\))]
The velocity which has 50\% of the integrated absorption on each side
and so acts as a figurative balancing point for the profile. This is
more accurate for irregular line shapes than the centroid of an
approximated Gaussian or Voigt profile, which will be misleading for an
asymmetrically shaped feature. Error bars are found through Monte
Carlo simulation by perturbing the measured profile with random errors
drawn from a Gaussian fit to the measured error distribution and
repeating the above computations through 100 iterations, after which the
standard deviation of the distribution of obtained values is reported as
the uncertainty.
\item[Velocity width (\(W_{90\%}\))]
The distance on the velocity axis from 5\% to 95\% integrated
absorption. This description of the line width is more robust to
irregular line shapes than the more common FWHM, yet also robust to
uncertainty determining the exact border between what is considered
continuum and what is considered absorption feature. It should, however,
be kept in mind when comparing to other results, that this width
encompases a larger fraction of the total velocity distribution than the
FWHM. Error bars found through the Monte Carlo method described above.
\item[Full Width at Half Maximum]
Standard FWHM in velocity space. This is found using the Python package
\texttt{statsmodels} to fit the observed profile to a Local Linear
Estimator (LLE) with a kernel width found through cross validation.
Error bars are found through Monte Carlo simulation. The FWHM is not
very sensitive to choice of kernel width.
\item[Velocity at 95\% integrated obsorption (\(v_{95\%}\))]
The absolute value of the velocity which has 95\% of the integrated
absorption on its red side. This velocity gives a tracer of the far end
of the velocity distribution relative to the systemic zero point.
\item[Velocity at minimum intensity (\(v_{min}\))]
The velocity at which the intensity of the line has its minimum. In a
perfectly symmetric absorption feature, this will be coincident with
\(v_{int}\).
\item[Ly\(\alpha\) peak velocity (\(v_{\rm peak}^{\rm Ly\alpha}\))]
For the galaxies that show Ly\(\alpha\) emission in the COS aperture, we
also report the velocity of the peak of this emission line. Like the
FWHM, this is found fitting it to an LLE with a kernel width found by
cross validation.
\end{description}

\begin{figure}[htbp]
\centering
\includegraphics{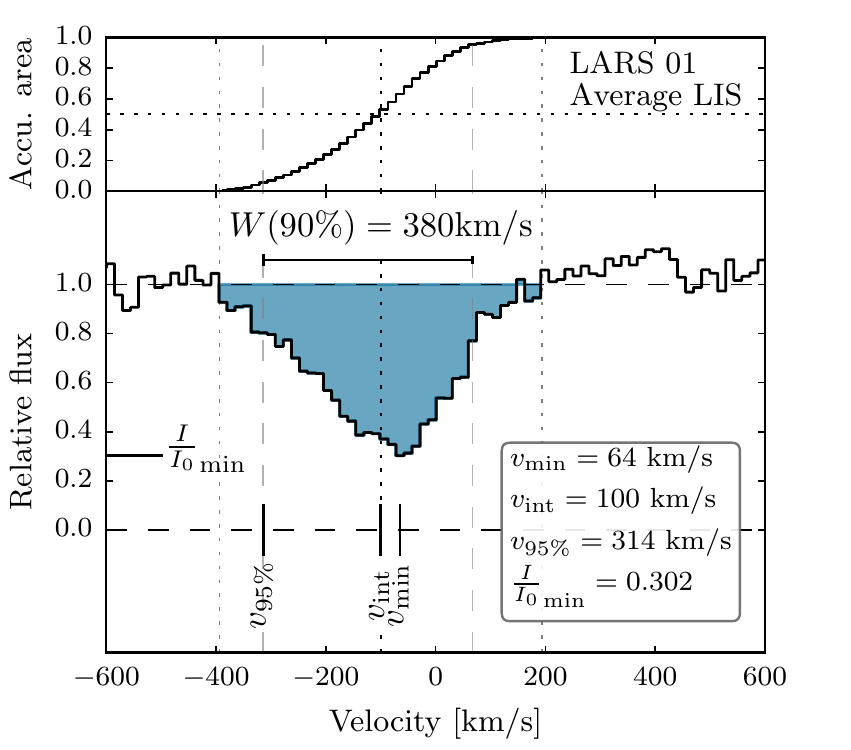}
\caption{Example of measurement of the characteristic velocities.
\textbf{Upper panel:} Accumulated absorption as a fraction of the total
area of absorption. \textbf{Lower panel:} Averaged low-ionization
absorption profile shown in black steps, with absorbed area shown in
blue (area above the level 1 counts as negative) and \(I/I_{0, \min}\)
and the characteristic velocities annotated. \textbf{Vertical dotted
lines} mark the center and limits of the absorption feature. \textbf{Vertical
dashed lines} mark the 5\% and 95\% levels of accumulated absorption.
\label{fig:dynvel}}
\end{figure}

Figure~\ref{fig:dynvel} shows an illustration of the quantities
mentioned above, except for FWHM which was measured separately and is
assumed to be well known. The averaged absorption profile is drawn in
black steps, the absorbed area in shaded green. The velocities
\(v_{\rm int}\), \(v_{\min}\), \(v_{95\%}\) and \(W_{90\%}\) are shown
along with the residual intensity at maximum absorption.

Besides these quantities, we have performed preliminary computations of
the continuum-subtracted Ly\(\alpha\) flux within the COS aperture to
evaluate how well the global Ly\(\alpha\) properties reflect the
Ly\(\alpha\) properties that arise as direct consequence of the
configuration and kinematic properties of the neutral gas.

An overview of the measurements described above is listed in
Tables~\ref{tab:results} and~\ref{tab:lyaprop}.

\section{Results}\label{sec:results}

\begin{deluxetable*}{clrlrcll} 
\tablecaption{Derived properties of the LARS galaxies.\label{tab:results}} 
\tablecolumns{8} 
\tablewidth{0pt} 
\tablehead{
% First row:
    \colhead{ID} & \colhead{$W_{90\%}$} &
    \colhead{$v_{\mathrm{int}}$} & \colhead{$v_{95\%}$} & \colhead{$v_{\min}$} &
    \colhead{$f_{C, \max}$} & \colhead{$\rm FWHM_{Si \textsc{ii}}$} & 
    \colhead{$\rm FWHM_{H\textsc{i}}$} \\
% Second row:
    \colhead{} &\colhead{[km s$^{-1}$]} & 
    \colhead{[km s$^{-1}$]} & \colhead{[km s$^{-1}$]} &\colhead{[km s$^{-1}$]} &
    \colhead{} & \colhead{[km s$^{-1}$]} &\colhead{[km s$^{-1}$]}\\
% Third row:
    \colhead{(1)} & \colhead{(2)} & \colhead{(3)} & \colhead{(4)} & 
    \colhead{(5)} & \colhead{(6)} & \colhead{(7)} & \colhead{(8)} 
    } 
\startdata 
LARS 01  & 381 \(\pm\) 34  & -100 \(\pm\) 12 & -314 \(\pm\) 24  & -65 \(\pm\) 21  & 0.70 \(\pm\) 0.03 & 270 \(\pm\) 33  & 160 \(\pm\) 10  \\ 
LARS 02  & 351 \(\pm\) 102 & -48 \(\pm\) 33  & -258 \(\pm\) 80  & -24 \(\pm\) 25  & 0.86 \(\pm\) 0.07 & 168 \(\pm\) 66  & 140 \(\pm\) 4\tablenotemark{*}\\ 
LARS 03  & 628 \(\pm\) 51  & -117 \(\pm\) 73 & -462 \(\pm\) 42  & -112\(\pm\) 53  & 0.97 \(\pm\) 0.05 & 517 \(\pm\) 22  & 170 \(\pm\) 20\tablenotemark{*}\\ 
LARS 04  & 365 \(\pm\) 61  & -31 \(\pm\) 152 & -261 \(\pm\) 46  & -13 \(\pm\) 31  & 0.99 \(\pm\) 0.07 & 311 \(\pm\) 42  & 180 \(\pm\) 4\tablenotemark{*}\\
LARS 05  & 466 \(\pm\) 47  & -138 \(\pm\) 14 & -390 \(\pm\) 22  & -84 \(\pm\) 24  & 0.81 \(\pm\) 0.03 & 347 \(\pm\) 17  & 160 \(\pm\) 28  \\ 
LARS 06  & 553 \(\pm\) 43  & 68 \(\pm\) 28   & -244 \(\pm\) 38  & 4 \(\pm\) 93    & 0.84 \(\pm\) 0.06 & 475 \(\pm\) 197 & 370 \(\pm\) 3\tablenotemark{$\ddagger$} \\ 
LARS 07  & 392 \(\pm\) 68  & -38 \(\pm\) 18  & -267 \(\pm\) 63  & 9 \(\pm\) 35    & 0.80 \(\pm\) 0.03 & 153 \(\pm\) 59  & 100 \(\pm\) 32  \\ 
LARS 08  & 522 \(\pm\) 85  & -240 \(\pm\) 55 & -442 \(\pm\) 70  & -225\(\pm\) 53  & 0.98 \(\pm\) 0.07 & 410 \(\pm\) 65  & 240 \(\pm\) 4\tablenotemark{*}  \\ 
LARS 09  & 503 \(\pm\) 34  & 40 \(\pm\) 140  & -263 \(\pm\) 27  & 82 \(\pm\) 50   & 1.00 \(\pm\) 0.05 & 463 \(\pm\) 19  & 310 \(\pm\) 18\tablenotemark{*} \\ 
LARS 10  & 484 \(\pm\) 51  & -27 \(\pm\) 187 & -287 \(\pm\) 38  & -3 \(\pm\) 49   & 1.03 \(\pm\) 0.08 & 446 \(\pm\) 27  & 280 \(\pm\) 140 \\ 
LARS 11  & 433 \(\pm\) 176 & -194 \(\pm\) 87 & -398 \(\pm\) 68  & -175\(\pm\) 56  & 1.04 \(\pm\) 0.09 & 517 \(\pm\) 184 & 260 \(\pm\) 11  \\ 
LARS 12  & 503 \(\pm\) 56  & -3 \(\pm\) 43   & -289 \(\pm\) 54  & -21 \(\pm\) 37  & 0.82 \(\pm\) 0.03 & 375 \(\pm\) 26  & \nodata         \\ 
LARS 13  & 484 \(\pm\) 43  & -105 \(\pm\) 22 & -359 \(\pm\) 36  & -85 \(\pm\) 105 & 0.76 \(\pm\) 0.04 & 373 \(\pm\) 52  & \nodata         \\ 
LARS 14  & 485 \(\pm\) 122 & -226 \(\pm\) 65 & -461 \(\pm\) 110 & -336\(\pm\) 66  & 0.40 \(\pm\) 0.05 & 129 \(\pm\) 59  & \nodata 
\enddata 
\tablenotetext{$\ddagger$}{From GMRT interferometry (J. M. Cannon et al, in
prep.)} 
\tablenotetext{*}{From VLA 21 cm interferometry. The remaining figures
are measured from GBT single-dish observations.} 
\tablecomments{Cols.  2-5 and 7 show the \textbf{line widths and} derived velocities described in
Sect.~\ref{sec:velocities}. Col. 6 shows the maximum velocity-binned covering
fractions. Col. 8 shows the FWHM of the HI lines as reported by \citet{LARSIII},
except $\ddagger$.} 
\end{deluxetable*}

\begin{deluxetable}{ccccc}
	\tablecaption{Ly$\alpha$ properties of the LARS
	galaxies.\label{tab:lyaprop}}
\label{tab:lyaprop}
\tablehead{
%First row
        \colhead{ID} & \colhead{$W_{\mathrm{Ly}\alpha}^{\mathrm{glob.}}$}&
        \colhead{$v_{\mathrm{peak}}^{\mathrm{Ly\alpha}}$} &
        \colhead{$f_{\mathrm{esc}}^{\mathrm{Ly}\alpha}$} & 
        \colhead{$f_{\mathrm{em, local}}^{\mathrm{Ly\alpha}}$} \\
% Second row
        \colhead{} & \colhead{[\AA]} & \colhead{[km s$^{-1}$]} & 
	\colhead{} & \\
% Third row
        \colhead{(1)} & \colhead{(2)} & \colhead{(3)} & \colhead{(4)} & 
	\colhead{(5)}}
\startdata{}
LARS 01 & 33.0 & 125 $\pm$ 7 & 0.119 & 0.028 \tabularnewline{}
LARS 02 & 81.7 & 150 $\pm$ 7 & 0.521 & 0.093 \tabularnewline{}
LARS 03 & 16.3 & 347 $\pm$ 8 & 0.003 & 0.000 \tabularnewline{}
LARS 04 & 0.00 & 443 $\pm$ 7 & 0.000 & 0.003 \tabularnewline{}
LARS 05 & 35.9 & 166 $\pm$ 7 & 0.174 & 0.032 \tabularnewline{}
LARS 06 & 0.00 & \nodata     & 0.000 & 0.000 \tabularnewline{}
LARS 07 & 40.9 & 178 $\pm$ 7 & 0.100 & 0.037 \tabularnewline{}
LARS 08 & 22.3 & 114 $\pm$ 7 & 0.025 & 0.005 \tabularnewline{}
LARS 09 & 3.31 & \nodata     & 0.007 & 0.000 \tabularnewline{}
LARS 10 & 8.90 & \nodata     & 0.026 & 0.000 \tabularnewline{}
LARS 11 & 7.38 & \nodata     & 0.036 & 0.000 \tabularnewline{}
LARS 12 & 8.49 & 415 $\pm$ 8 & 0.009 & 0.006 \tabularnewline{}
LARS 13 & 6.06 & 276 $\pm$ 9 & 0.010 & 0.003 \tabularnewline{}
LARS 14 & 39.4 & 254 $\pm$ 8 & 0.163 & 0.119 
\enddata{}

\tablecomments{Col.~2 shows global Ly$\alpha$ equivalent width, 
  Col.~3 shows (where present) the Ly$\alpha$ peak velocity computed as 
  described in Sect.~\ref{sec:velocities}, Col.~4 lists global, imaging-derived 
  Ly$\alpha$ escape fractions; Col.~5 lists fraction of intrinsic Ly$\alpha$ 
  emitted into the COS aperture. Values of Cols.~2 and 4 were first reported in 
  \citet{LARSII}.}
\end{deluxetable}

\subsection{Line profiles}\label{line-profiles}

\subsubsection{Low-ionized state}\label{low-ionized-state}

LIS absorption profiles, as  shown in
Figure~\ref{fig:liproplot}, show a variety of shapes, depths,
complexity and blueshift compared to the nebular zero point. Line widths span
almost a factor of two, with FWHM ranging from \(\rm \sim 350\) km s\(^{-1}\)
as the lowest and \(\sim 630\) km s\(^{-1}\) as the highest.

In most of the galaxies, all four Si {\sc ii} profiles show almost identical
shape and depth. There are exceptions; most notably, Si {\sc ii} \(\lambda\)
1304 is somewhat shallower than the other transitions in LARS 01, 05,
07, 11, 12 and 13. Likewise, but less notably, Si {\sc ii} \(\lambda\) 1260 in
several cases dips \emph{below} the other lines, especially in the blue
wing, as seen most clearly in e.g.~LARS 03, 04, 09 and 10. But
considering the differences in transition strengths \(f_{ik}\), which
are up to a factor of ten, the lines have very similar shapes and
depths, in some cases practically identical. This is usually a sign that
the system is fully optically thick in all the involved transitions, with
the cases in which there is residual emission at maximum absorption
being due to partial covering. However, \citet{Prochaska2011} and
\citet{Scarlata2015} have shown that this effect can in some
cases be mimicked by the effect of scattered light re-emitted into the
line of sight; see Section~\ref{sec:discuss}.

Yet other possible causes of residual intensity at line bottom exist.
For one, the systems can be part transparent; but this usually will show
stronger difference in line strengths for different transitions of the
same species, depending on \(f_{ik}\) for each transition.

Residual emission can also be due to resolution effects. The COS
aperture is centered on the strongest star-forming knot of each galaxy.
As seen in Table~\ref{tab:results} col. 2, these knots have an effective
resolution quite close to the COS resolution of \(\sim 20,000\).
But the less luminous parts of the galaxies generally give some emission
across the aperture. Spatially extended emission within the aperture is
more poorly resolved, and so emission from neighboring velocity ranges
may bleed into line center and give a false impression of a perforated
medium. Judging from the aperture maps, this could mainly be a concern
for LARS 3, 4, 5, 8, 9, 11 and 12, and possibly 10. Of these, LARS 3, 4,
8, 9, 10 and 11 have no or almost no residual intensity at maximum
absorption, in which cases this effect evidently is not strong enough to
cause false impressions of a perforated medium. LARS 5 and 12 have quite
flat profiles around line minimum, implying that this effect should be
quite strong to account for the residual intensity at line center. The
uncertainties caused by the effect of reemission in the absorption
trough described above are likely to dominate over this one in these
cases.

Looking at Figure~\ref{fig:liproplot}, two other tendencies stand out.
For one, Si {\sc ii} 1304 is in some cases more shallow than the other lines.
This is often, though not consistently, in conjunction with a generally
larger spread in the depths of the Si {\sc ii} profiles. This could possibly
be of interest to the discussion of Section~\ref{subs:reemission}.

The other trend is the profile of C II \(\lambda\) 1334 generally having
a tendency to be deeper than the other lines; in particular, there seems
to be what could be a weaker, contaminating line in the red wing. This
could possibly be due to the higher ionization potential of C II (see
Table~\ref{tab:lines}), meaning it could survive not only in the warm
neutral medium but also in a warmer state, which could add to its
absorption feature. This would affect the averaged profile by making it
deeper in this velocity range than actually warranted by the strength of
C II 1334. However, as is seen on the figure, the emission of O {\sc i}*
\(\lambda\) 1305, where present, tends to counteract this in the
computation of the averaged profile, and in general, being one out of
six lines included, it only has a slight impact on the averaged profile.
One should however keep in mind that it could have a slight impact on
quantities like \(v_{\rm int}\), which would be artificially lowered by
this, and on \(v_{95\%}\) which could be slightly elevated. However,
this is most likely negligible compared to the uncertainties due to
measurements errors as listed in Table~\ref{tab:results}.

The line of Si {\sc ii} 1304 often shows some blending in its red wing with
the fluorescent emission companion at \(\lambda\) 1305 Å to the O {\sc i} 1302
line. In most cases this only has a slight effect on the far red wing of
the profile, but as seen in Figure~\ref{fig:liproplot}, it could be
responsible for the red wing discrepancies between the \(\lambda\) 1304
Å and the other Si {\sc ii} lines in LARS 3, 4, 7 and 12.

\subsubsection{High-ionized state}\label{high-ionized-state}

The averaged profile of the three Si {\sc iv} lines found in our spectra (or the
available subset hereof) was found for each galaxy and are shown in
Figure~\ref{fig:phases} together with the stacked LIS profile. The majority of
Si {\sc iv} is expected to be found in gas from hot winds from the regions
surrounding the starbursts, ramming into the neutral medium, which is
heated and then gradually cools again, while driving the outflow \citep[see
e.g.][]{Grimes2009, Strickland2009}. The Si {\sc iv} profiles are therefore
expected to more or less follow the morphology of the LIS profile, although with
slightly higher outward velocity. Looking at the plots, this is evidently the
case for many galaxies, like e.g.~LARS 2, 4, 6, and 7; but the effect is less
obviously present in others.

An in-depth analysis of the Si {\sc iv} profiles is outside the scope of this
work, but it is interesting to note that 1) The red wing of the LIS
profile in LARS 9 is significantly deeper than the Si {\sc iv} profile, and 2)
that LARS 14 has a significantly deeper Si {\sc iv} than LIS profile, which
confirms that this galaxy is highly ionized.

The spectrum of LARS 11 has no available Si {\sc iv} features,
they all fell outside the detector range except \(\lambda 1122\) which was
coincident with geocoronal Ly\(\alpha\).

\subsection{Neutral gas covering}\label{neutral-gas-covering}

\begin{figure*}[htbp]
\centering
\includegraphics{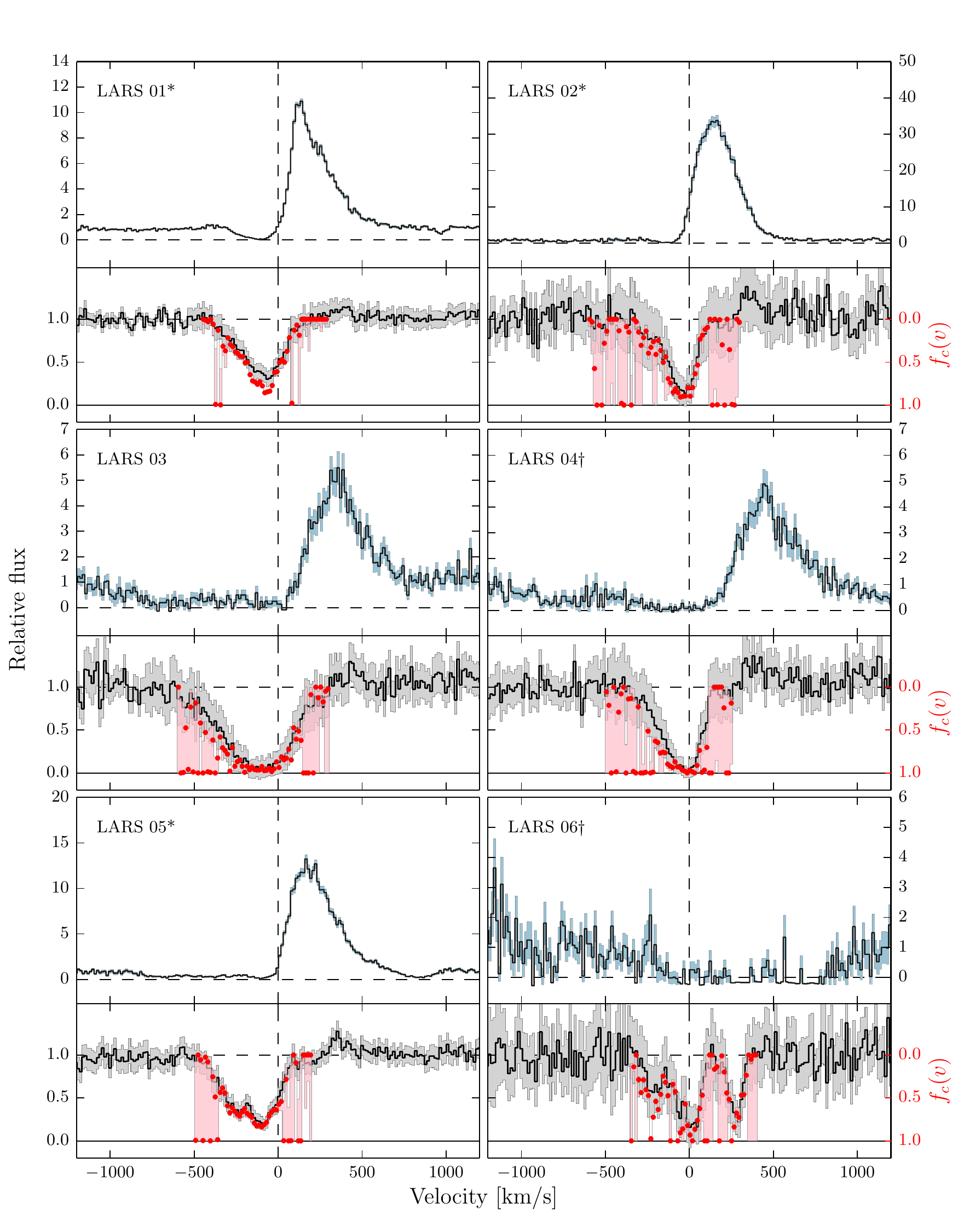}
\caption{\textbf{Upper panels:} Ly\(\alpha\) profiles of LARS 1 through
6 with \(\pm 1\sigma\) regions shown as blue shading above and below.
\textbf{Lower panels:} Covering fractions of neutral gas for the same
galaxies shown as red dots with light red shading indicating confidence
intervals calculated by the method described in Section~\ref{sec:covfrac}. Black
steps show the averaged low-ionized metal line profiles, with standard
errors indicated by gray shading. \label{fig:covfracs}}
\end{figure*}

\begin{figure*}[htbp]
\centering
\includegraphics{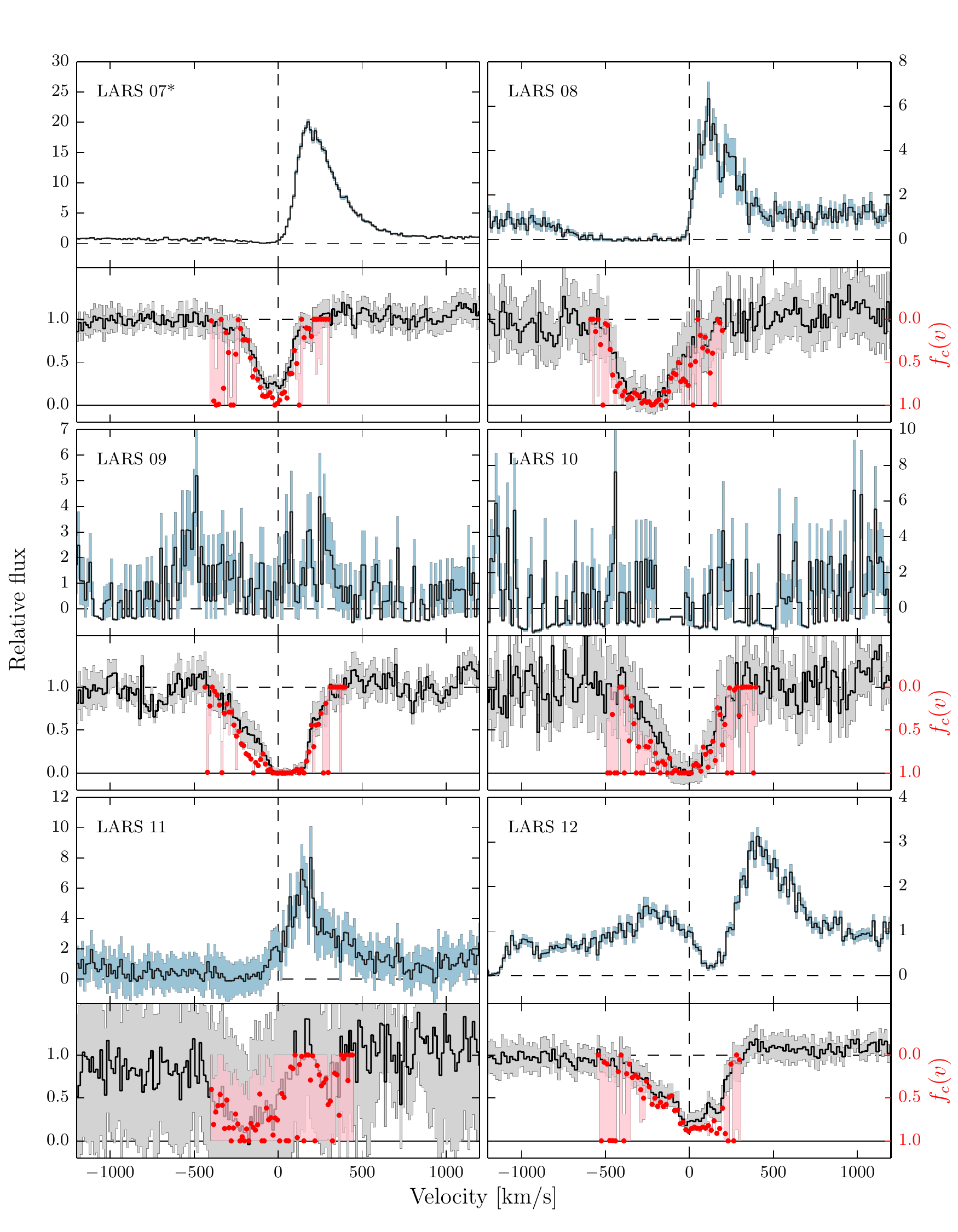}
\caption{Same as Fig.~\ref{fig:covfracs}, but for LARS 7 through 12.
\label{fig:coverfracs2}}
\end{figure*}

\begin{figure*}[htbp]
\centering
\includegraphics{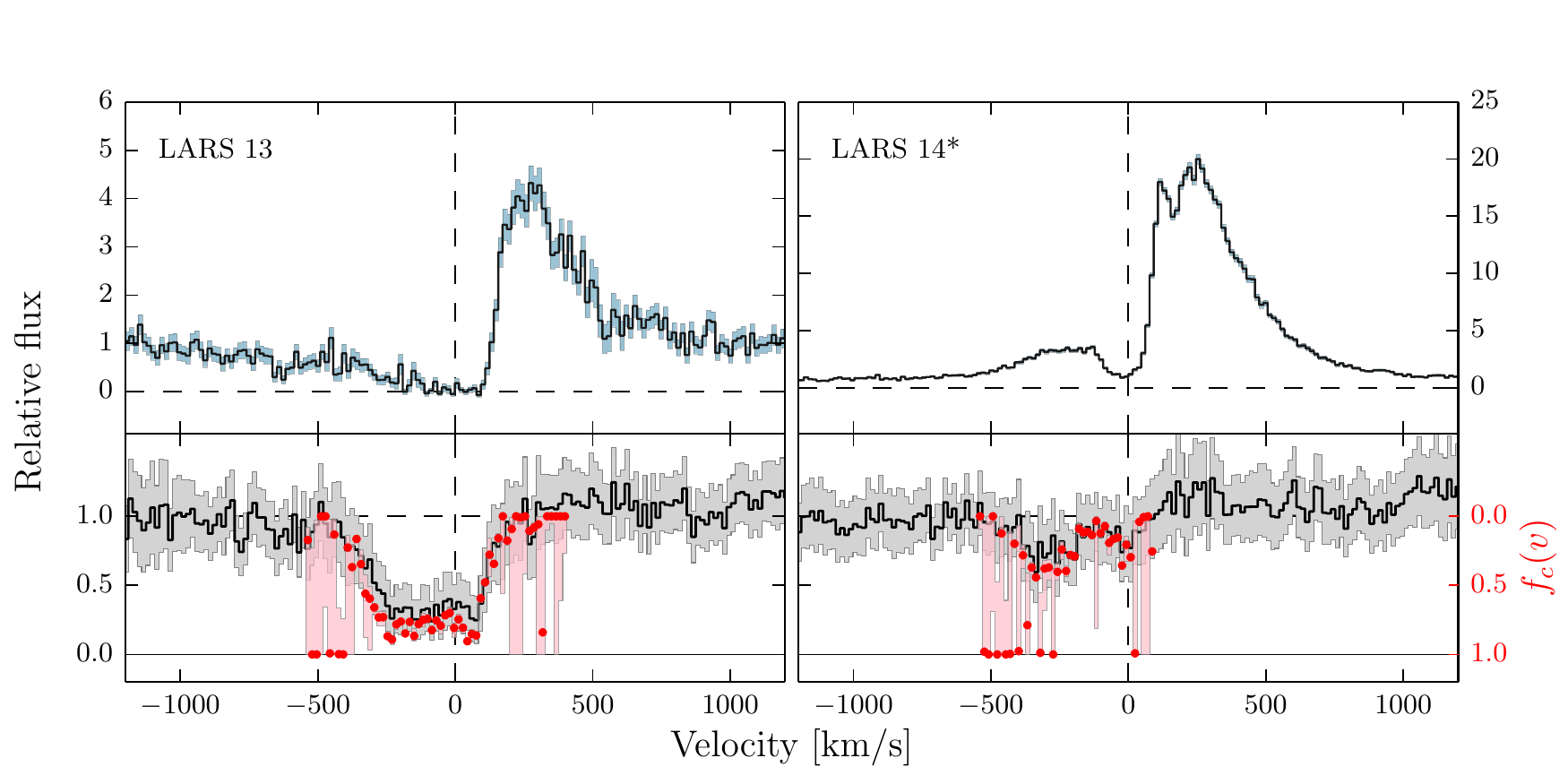}
\caption{Same as Fig.~\ref{fig:covfracs}, but for LARS 13 through 14.
\label{fig:coverfracs3}}
\end{figure*}

Neutral gas covering in narrow velocity ranges were computed for all 14
LARS galaxies based on the method described in Section~\ref{sec:covfrac}. The
results are visualized in Figures~\ref{fig:covfracs},~\ref{fig:coverfracs2} and
~\ref{fig:coverfracs3}, with black steps showing the averaged LIS profile
and propagated  one \(\sigma\) errorbands in shaded gray. Red dots
indicate, on a reversed scale marked in red on the right, the computed
neutral covering fractions for each velocity step. The pink shades show
confidence levels of the covering fractions, cut off at zero and one as
subzero and above-one fractions are unphysical.

We find that while there is a certain scatter in the computed covering
fractions, they generally coincide very well with \((1 - I/I_0)\) of the
averaged low ionized absorption profile. In some galaxies, there is a
tendency for the covering fraction plots to dip a bit below the stacked
profiles, especially in the velocity ranges close to maximum absorption.
These ranges are generally coincident with ranges in which either the Si
II \(\lambda 1304\) transition profile lying visibly above the other
lines, or the Si {\sc ii} \(\lambda 1260\) line profile lying significantly
below the others. If these are due to polluting features or other
systematics, this would cause the stacked profile to slightly
misrepresent the actual physical system, and could also cause minor
systematics in the fits for \(f_C(v)\). This effect introduces
uncertainties about the presence of residual emission at maximum
absorption for LARS 7 only; in general the assumption of complete
optical thickness seems well justified. We therefore from this point
have adopted \((1 - I/I_0)\) as a proxy for the velocity-binned covering
fractions, and we shall report the maximum fractional depth of the
profile as the maximum velocity-binned covering factor, \(f_{C, \max}\)
(see e.g.~Table~\ref{tab:results}).

About half of these stacked profiles are saturated in absorption, while
the rest show some residual intensity. Eight of these galaxies: LARS 01,
02, 05, 06, 07, 12, 13 \& 14, show residual intensity at maximum
absorption. All the strong Ly\(\alpha\) emitters from \citep{LARSII} -
LARS 01, 02, 05, 07 and 14 - are found within this group.

Three galaxies - LARS 6, 9 and 10 - show deep, damped Ly\(\alpha\)
absorption features in the aperture. Their LIS lines have in common a
relatively deep, static component of the neutral medium, as discussed in
the introduction. This component is completely black at line center for
LARS 9 and 10, but not for LARS 6. Interestingly, only LARS 6 is also a
global absorber in Ly\(\alpha\). This points at LARS 9 and 10 emitting
the majority of their Ly\(\alpha\) in extended halos, consistent with
what is visible in Figure~\ref{fig:apertures}. LARS 4, which is a global
absorber, shows some weak emission in the aperture, although not enough
to outweigh what is absorbed in the outer regions.

\begin{figure}[htbp]
\centering
\includegraphics{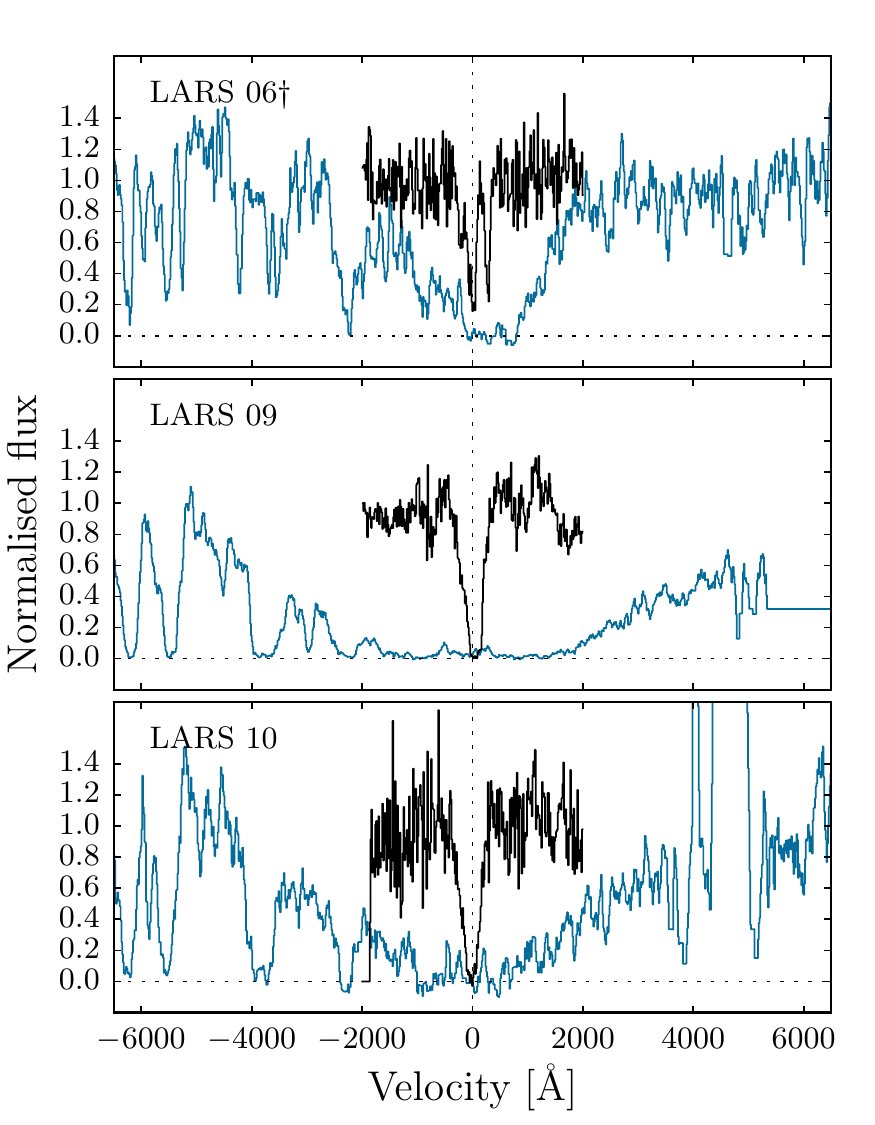}
\caption{Continuum-normalised Ly\(\alpha\) profiles of the three strong
absorbers of the sample, LARS 6, 9 and 10, shown in blue. The averaged,
normalized LIS profiles are shown in black for comparison.
\label{fig:broadlines}}
\end{figure}

The average LIS profiles and the Ly\(\alpha\) profiles of LARS 6, 9 and
10 are shown for comparison in Figure~\ref{fig:broadlines}. Especially
LARS 9 and 10 show a quite striking contrast in width between the metal
and the H {\sc i} profiles.

\subsection{Local and large scale Ly\(\alpha\)
transfer}\label{sec:locallya}

One major objective of this work is to compare detailed knowledge about
conditions close to the hot, star-forming regions that give rise to the
large majority of intrinsic Ly\(\alpha\) photons, obtained through SDSS
and COS spectroscopy, with the strength and morphology of global
Ly\(\alpha\) output, which is believed to depend strongly on these
central conditions. However, an important requirement for this to bear
any physical significance is that the conditions of these sight lines be
representative of general conditions in the galaxies studied. As a test
of this, we have computed the Ly\(\alpha\) flux emitted into the
aperture as a fraction of the intrinsic Ly\(\alpha\) emission in the
aperture, inferred from the above fluxes and H\(\alpha\) measured from
SDSS spectroscopy \citep{LARSI}.

Radiative transfer of Ly\(\alpha\) is, due to its strong resonance, a global
phenomenon for a galaxy. Photons interacting along the line of sight, seemingly
absorbed, may escape the galaxy eventually, while photons originating from
regions outside the aperture may scatter into it. We define the \emph{local
emitted Lyman alpha fraction}, \(f_{\rm em, local}^{\rm Ly\alpha}\) as the
Ly\(\alpha\) flux emitted into the aperture in units of the intrinsic
Ly\(\alpha\) flux within the aperture; this is listed in Col.~5 of
Table~\ref{tab:lyaprop}. This is not exactly an escape fraction, but the closest
comparable quantity that can be meaningfully defined in a local region of a
galaxy.

\begin{figure}[htbp]
\centering
\includegraphics{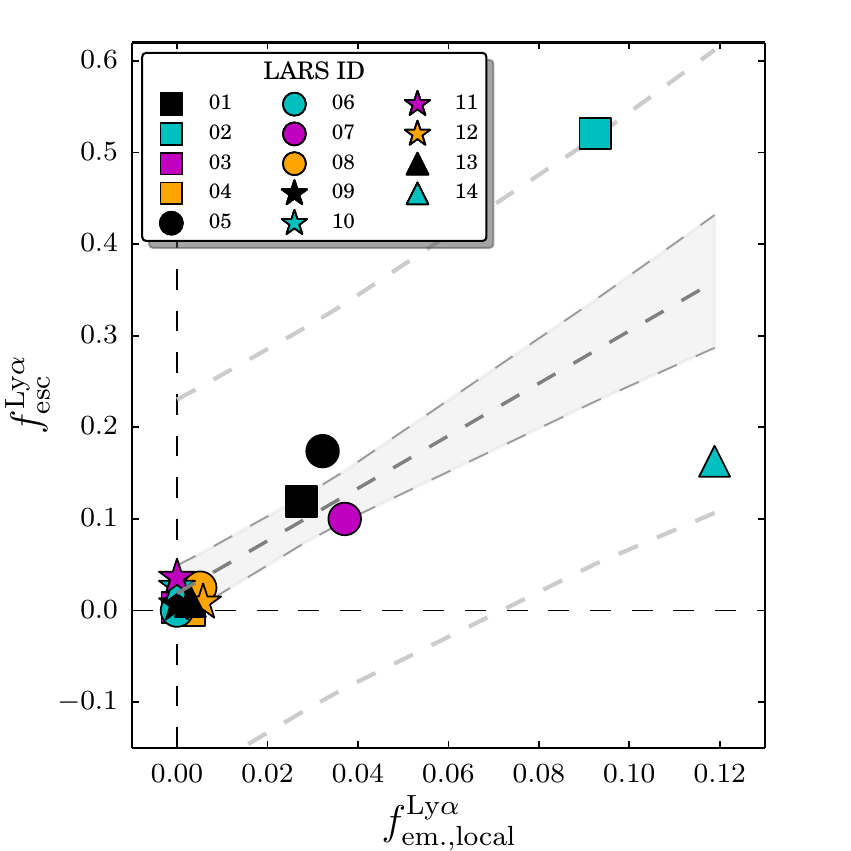}
\caption{Local emitted fraction vs.~global escape fractions of
Ly\(\alpha\) in the LARS galaxies. Also shown is the best linear fit
with the 68\% confidence band in shaded gray, and further out the 95\%
prediction band in light dashed gray. \label{fig:EscFracs}}
\end{figure}

Figure~\ref{fig:EscFracs} sees these locally emitted fractions plotted
against the global Ly\(\alpha\) escape fractions from \citet{LARSII}.
The data show a clear correlation, with LARS 2 and 14 as a outliers, 14
lying just outside the 95\% prediction band while LARS 2 is falling
somewhere between the 68\% and 95\% curve (see discussion below). LARS
14 shows a noticeably stronger local emission relative to the global
output than the remainder of the sample, while on the other hand LARS 2
has a stronger global and weaker local Ly\(\alpha\) output than usual.
The full sample shows a Pearson-\(r\) for these two quantities of
\(\sim 0.77\). Evidently, the aperture regions are well representing the
Ly\(\alpha\) RT conditions in the galaxies of the LARS sample. This also
means that in all subsequent figures showing relations involving
\(f_{\mathrm{esc}}^{\mathrm{Ly}\alpha}\), the in-aperture version of the
same would look essentially the same, except for LARS 2 and 14 which
would have lower and higher local escape, respectively.

\subsection{Importance of outflows}\label{importance-of-outflows}

\subsubsection{Lyman \(\alpha\) escape}\label{lyman-alpha-escape}

Previous studies of nearby galaxies \citep{Kunth98, Wofford2013}
conclude that an outflowing bulk motion of the neutral medium is
necessary for Ly\(\alpha\) to escape: \citet{Kunth98} analyzed
observations of 8 galaxies and found velocity offsets of up to 200 km s$^{-1}$
between O {\sc i}/Si {\sc ii} absorption lines and emission from the hot, ionized
gas in the emitting galaxies, while galaxies with a (near-) static ISM
showed deep, damped absorption features in Ly\(\alpha\).

\begin{figure}[htbp]
\centering
\includegraphics{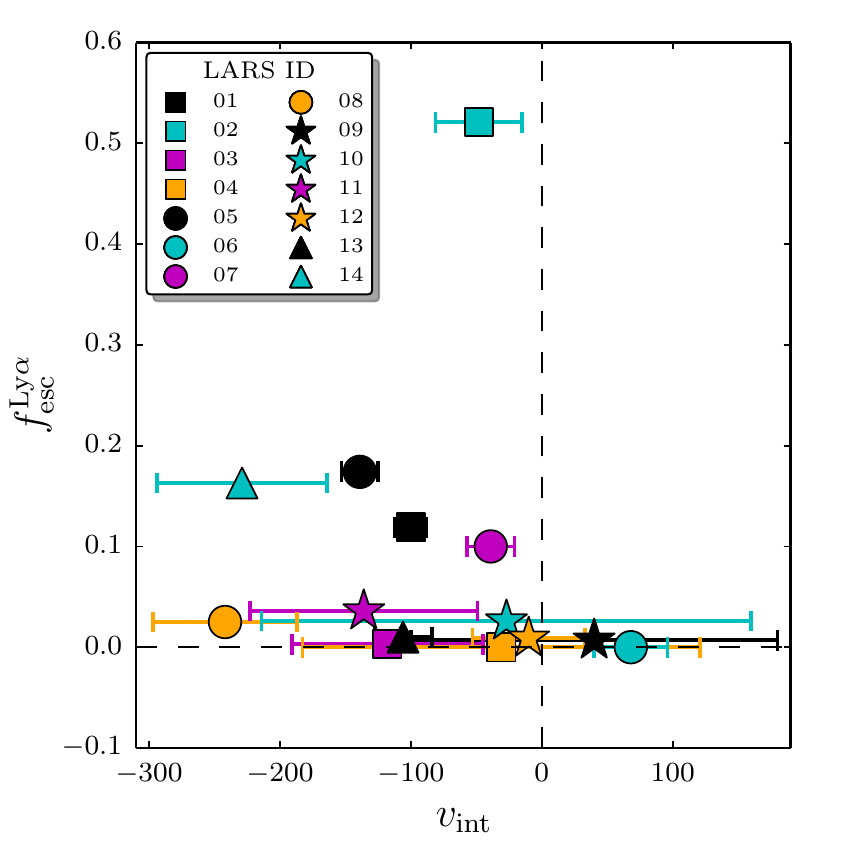}
\caption{Plot of wind velocity vs.~escape fraction
\(f_{\rm esc}^{\rm Ly\alpha}\) for the LARS sample. Escape fractions are
global, imaging-derived values from \citet{LARSII}.
\label{fig:WindEscFrac}}
\end{figure}

Our findings from the LARS sample are consistent with this picture (see
Table~\ref{tab:results}). Like \citet{Wofford2013} we also stress,
however, that other mechanisms must be in play; with a Pearson's
\(r = -0.1\), there is no statistical correlation between
\(f_{esc}^{Ly\alpha}\) and \(v_{int}\) in general, and we do see cases
of galaxies with outflowing winds and yet almost no emission in
Ly\(\alpha\), although no emission through a static medium is observed.

Outflowing winds must be present to allow for Ly\(\alpha\) escape, but
as can be seen in Figure~\ref{fig:WindEscFrac}, it is not sufficient to
guarantee this: LARS 03, 08, 11 and 13 show significant outflow
velocities, but still very low Ly\(\alpha\) escape fractions. Evidently,
other, competing effects must have significance.

\subsubsection{Lyman alpha peak
velocity}\label{lyman-alpha-peak-velocity}

From current models of Ly\(\alpha\) radiative transfer
\citep[e.g.][]{Verhamme2006}, it is predicted that in an environment of a
spherically symmetric outflowing medium of high H {\sc i} column density, the
emission feature will be dominated by a redshifted peak with a maximum
around \(v_{\rm peak}^{\rm \alpha} \sim 2 \times v_{\rm int}\), a
component which is backscattered on the far, inner, receding surface of
the expanding shell, while the component emitted directly towards the
aperture in these models is predicted to be largely suppressed while
traversing the medium. According to these models, we would expect to see
a correlation between outflow speed and Ly\(\alpha\) emission peak
velocity.

\begin{figure}[htbp]
\centering
\includegraphics{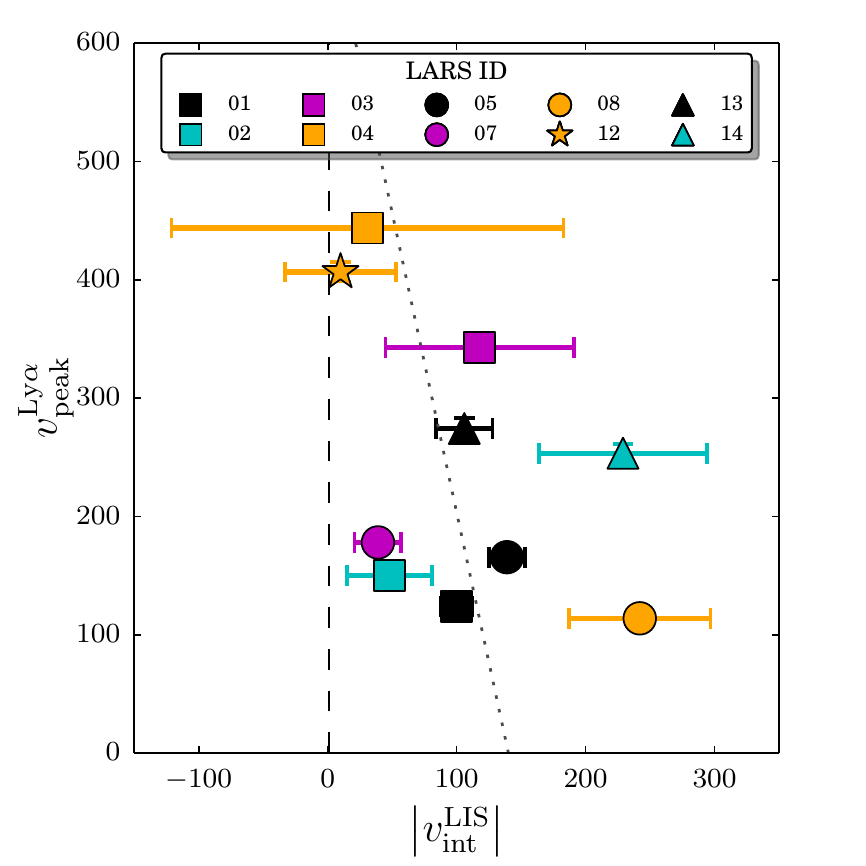}
\caption{Ly\(\alpha\) peak velocity vs.~wind speed for the subsample
that shows Ly\(\alpha\) emission in the COS aperture. The dotted line
shows the result of linear regression weighted by the uncertainties in
\(v_{\rm int}\).\label{fig:vlyavint}}
\end{figure}

It is therefore a bit surprising that we find no such correlation in our
sample. In Figure~\ref{fig:vlyavint}, these quantities are plotted
against each other for the 10 spectra in LARS with Ly\(\alpha\)
emission. In the figure is also shown, in dotted dark gray, the least
squares regression line weighted by the uncertainties in
\(v_{\rm int}\). Uncertainties on \(v_{\rm peak}^{\rm Ly\alpha}\) are
negligible in comparison and were not included in the fit.

The figure clearly shows that the expected correlation is not there.
There is in fact a weak anticorrelation between these velocities; the
direct opposite of the expected result, although with a Pearson-
\(r \approx 0.4\) it is not a strong correlation and should be taken
with a grain of salt. What we observe is more akin to the simpler model
discussed in e.g. \citet{MasHesse03} of an emission feature overlaid by
a multiplicative absorption profile.

\subsection{Effect of \(f_C\) on Ly\(\alpha\) escape
fraction}\label{effect-of-fux5fc-on-lyalpha-escape-fraction}

One should be careful when interpreting the covering fractions presented
in Sect.~\ref{sec:covfrac}. The measured maximum velocity-binned covering
fraction $f_C(v_{\mathrm{min}})$ (see e.g. Fig.~\ref{fig:dynvel}) constitutes
only a lower limit to the total covering fraction $F_C$ \citep{Jones2013}: A
system at velocity other than $v_{\mathrm{min}}$ may cover a projected area with
only partial or no overlap with the area covered by gas at $v_{\mathrm{min}}$.
Such a system will add to the total covering fraction, increasing it beyond the
minimum set by $f_C(v_{\mathrm{min}})$ - unless, of course,
$f_C(v_{\mathrm{min}})$ is unity.

\begin{figure}[htbp]
\centering
\includegraphics{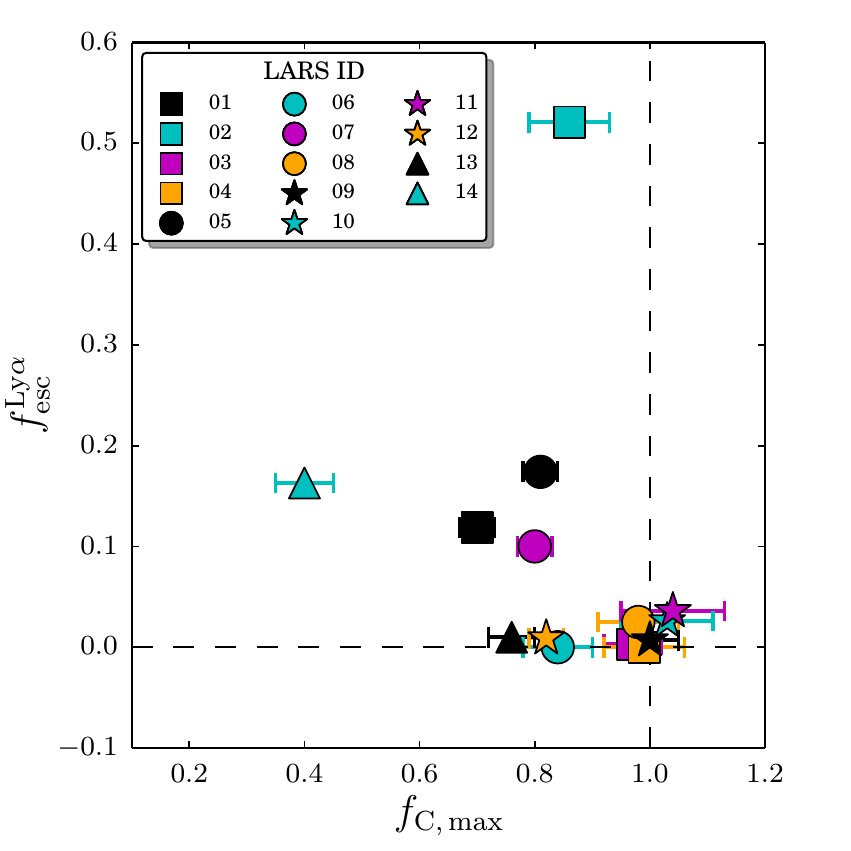}
\caption{The effect of covering fraction on Ly\(\alpha\) escape
fraction. When the maximum covering fraction is unity, no Ly\(\alpha\)
photons escape. When the maximum covering fraction is significantly
below unity, Ly\(\alpha\) photons can escape. Colors and symbols as
above. \label{fig:fesc-cofra}}
\end{figure}

Figure~\ref{fig:fesc-cofra} shows a plot of Ly\(\alpha\) escape fraction
vs.~maximum velocity-binned covering fraction \(f_{C, \max}\) as
described above. A negative residual intensity is of course unphysical
and would be interpreted as zero. It is immediately clear that at a
covering fraction of 1, very little Ly\(\alpha\) radiation escapes. This
physically corresponds to a full sheet of neutral gas screening off the
emitting star clusters. All Ly\(\alpha\) emitters of the sample have
\(f_{C} \lesssim 0.9\).

\subsection{Star formation and perforation: a possible
clue}\label{sec:ewha-fc}

\begin{figure}[htbp]
\centering
\includegraphics{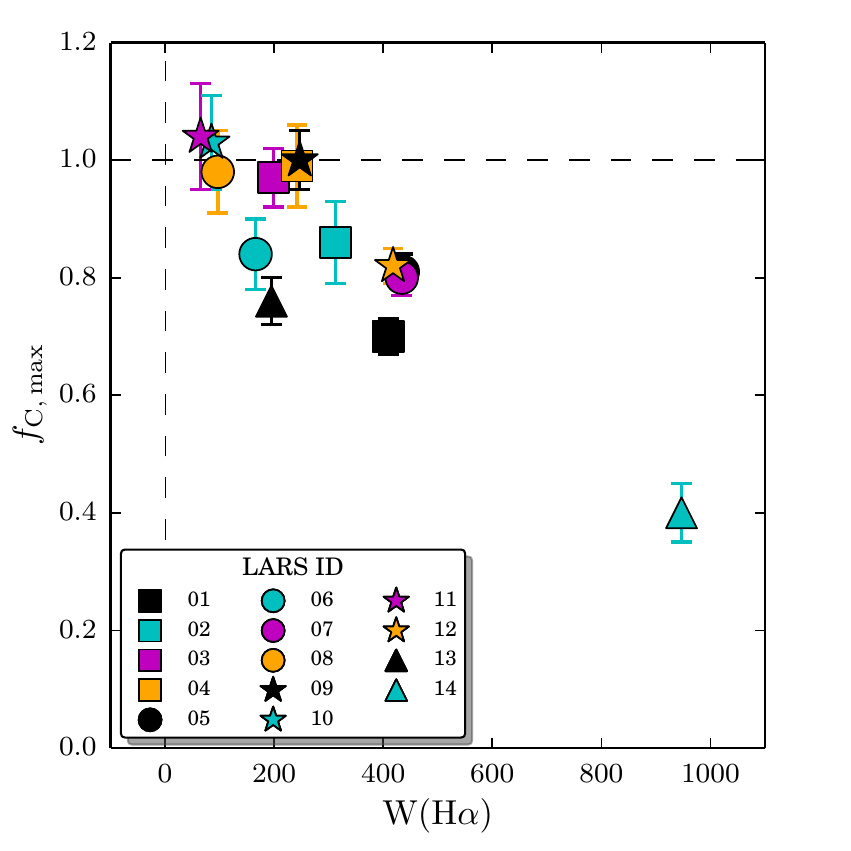}
\caption{Maximum covering fraction vs.~global \(\rm H\alpha\) equivalent
width. We find a clear anticorrelation between the two quantities.
\label{fig:MoneyShot}}
\end{figure}

One quite surprising finding of this work came from comparing Figure~\ref{fig:fesc-cofra} to Figure~5 in \citet{LARSII}, panel 3 in the lower
row, comparing W(H\(\alpha\)) to \(f_{\rm esc}^{\rm Ly\alpha}\). The
similarity is quite striking, pointing at a possible relation between
EW(H\(\alpha\)) and \(f_{C, \max}\). Figure~\ref{fig:MoneyShot} shows
that this is indeed the case, the two are - with all due reservations
regarding sample size - strongly correlated. This is, to our knowledge,
the first time such a correlation has been observed.

\subsection{Interpretation}\label{sec:interp}

\subsubsection{Kinematics and geometry}\label{kinematics-and-geometry}

Our results suggest a picture of a neutral medium consisting of multiple
dense clouds flowing predominantly outwards at a range of velocities,
although also contributions redwards of line center are found in some of
our sample galaxies. Considering the width and morphology of the line
profiles, it seems clear that they consist of multiple contributions
from separate subsystems which move outwards at a range of velocities
and therefore must be physically disjoint. These clumps are generally
optically thick near the centers of the Si {\sc ii} lines, but each may only
partially cover the starburst region and hence leave residual flux even
at the central wavelength of the absorption line for each subsystem.
Together, the absorption features will blend into one line of
complicated morphology, as observed. At each velocity, a part of the
starburst is covered by a completely opaque cloud, while others are
unblocked.

Consider a toy system of two clouds which each cover 50\% of the
starburst. If these move at the same velocity, they will block all light
shifted into the line center in that frame. If, however, they have a
velocity difference large enough to completely deblend the line
contributions, the resulting profile would show two separate dips in
flux density down to 50\%, while at least 50\% of photons at any given
wavelength would encounter no opacity at all when traversing the system.

\begin{figure*}[htbp]
\centering
\includegraphics{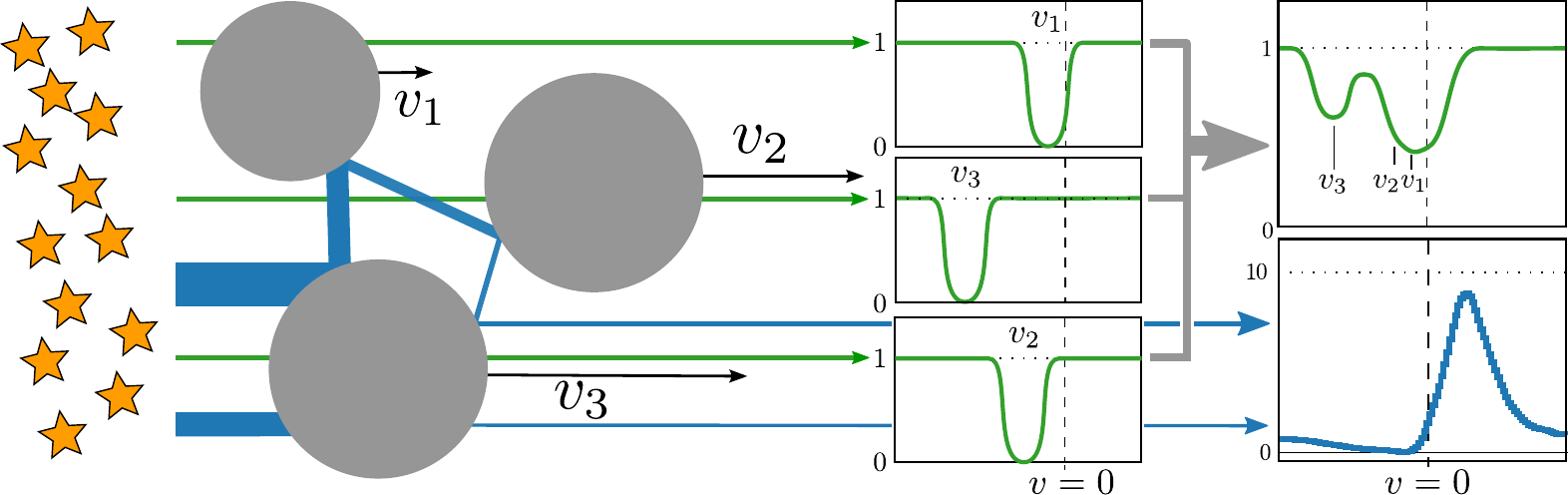}
\caption{Outflowing wind consisting of clumps at different velocoties
which each cover the starburst only partly, while together covering
completely or almost completely. Green arrows show UV continuum
emisssion in accordance with the color coding of Figure~\ref{fig:apertures}. Metal transitions assumed to be optically thick
leave saturated absorption features in the light passing through, shown
in each of the three small middle panels. None of the clouds are
completely covering the star-forming regions and hence, when integrated
over the aperture, together they leave a complex line profile as shown
in the large panel to the right. Ly\(\alpha\) radiation is shown in
blue, likewise in accordance with the coloring scheme of the three-color
aperture map images. Ly\(\alpha\) can escape in small quantities through
the clumps but can also, like UV continuum (not shown on this figure),
scatter on dust grains and thus, in the presence of some pathway through
a medium with a projected covering fraction of 1, undergo much fewer
scatterings and thus have a strongly enhanced chance of escape. (see
e.g. \citet{Duval2014}). \label{fig:outflow}}
\end{figure*}

The model is illustrated in Figure~\ref{fig:outflow}, which should be
read as a radial beam cut out of a spherically symmetric system. The
stars to the left signify the background sources, including the H {\sc ii}
regions in which the recombination lines like Ly\(\alpha\) are emitted.
Next to these, optically thick clumps of the neutral medium are shown as
gray disks, moving outwards at different velocities. FUV continuum
shining through the clumps is shown as green arrows and Ly\(\alpha\)
radiation is shown in blue, in accordance with the coloring scheme in
Fig.~\ref{fig:apertures}. Fractions of the FUV pass through the clouds,
which each leave saturated absorption features in parts of the light. If
the difference in velocities surpasses the line widths of the individual
systems, the absorption feature of each clump will fall at different
wavelengths in the spectrum, and the line will show residual flux
everywhere, even though it consists of overlapping, saturated clumps which
combined cover the source completely. The
three green absorption profiles shown in the middle are the line
imprints left by the individual clumps, and the larger green profile in
the upper right shows the resulting profile when the light from the
three different projected regions is integrated over the aperture: the
simple, saturated contributions of each clump can, even when the clumps
are together fully covering the background source, add up to a quite
complex line profile with residual flux everywhere. In the lower right,
in blue, is shown a typical P Cygni-shaped Ly\(\alpha\) profile as it
would often look from a system with unity combined covering fraction
like this one.

\begin{figure*}[htbp]
\centering
\includegraphics{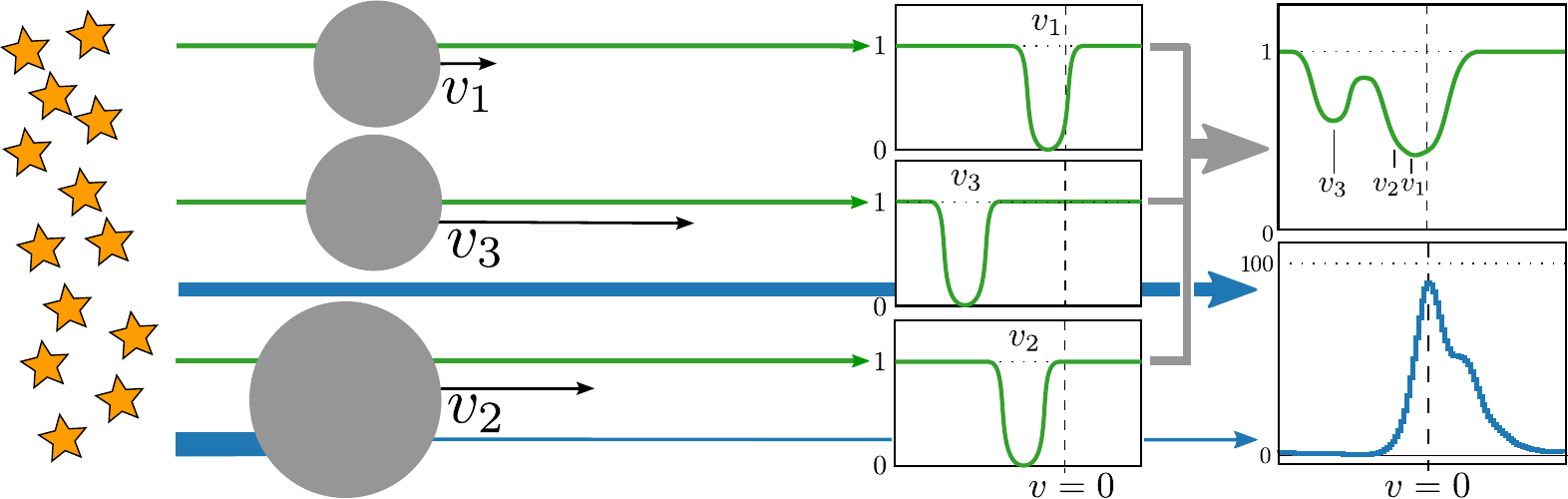}
\caption{Like Figure~\ref{fig:outflow}, but this time with a projected
covering fraction \(F_C < 1\), the ``picket fence'' model of
\citet{Heckman2011}. In this case, even in the presence of few very
narrow direct sight lines to the star forming regions can make these
freely escaping Ly\(\alpha\) photons dominate the shape of the line
profile (\citet{Duval2014}). In any case, any significant direct escape
means that the Ly\(\alpha\) profile never goes completely black, whereas
in the case of full covering, all Ly\(\alpha\) photons will undergo
scattering and the line profile thus become very similar to that of a
simple screen model, with its characteristic dark, deep absorption
around line center. The metal absorption lines in the UV continuum may
however be practically indistinguishable. \label{fig:picketfence}}
\end{figure*}

In Figure~\ref{fig:picketfence} we show for comparison a sketch of the ``Picket
Fence'' model of \citet{Heckman2011}. It is very similar to
Figure~\ref{fig:outflow}, except the clumps are now only partly covering,
leaving holes in the neutral medium allowing for direct escape of Ly\(\alpha\)
and hence also LyC. The individual clumps still leave saturated absorption lines
in parts of the FUV light, which add up to a more complex line profile with
residual intensity everywhere. Here, the residual intensity is only in part due
to the velocity offsets between the clumps, while a contribution comes directly
through holes in the neutral medium. The combined metal absorption feature of
such a system is difficult to distinguish from the one seen in the \(F_C = 1\)
case above. The Ly\(\alpha\) feature will, however, have a nonzero flux in the
line center and, depending on the size of the holes, maybe even be dominated by
the narrow, undistorted component directly leaking through.  We therefore argue
that a well resolved Ly\(\alpha\) spectrum and precise systemic velocity zero
points are necessary in order to distinguish between these two similar, but not
identical cases.

\subsubsection{ISM evolution}\label{ism-evolution}

By combining column 9 in Table~1 of \citet{LARSII} with columns 5 and 11
in Table~2 in same, we find that \(W(\rm H\alpha)\) is well correlated
with SSFR with a Pearson-\(r = 0.77\), so that \(W(\rm H\alpha)\) is a
good proxy for specific star formation. This suggests that star
formation feedback could be driving the fragmentation and perforation of
the neutral medium; rather than uniformly pushing the LIS medium
outward, the feedback would cause growing Rayleigh-Taylor instabilities
in the outflowing medium, which would fragment into clumps of different
sizes and velocities \citep{MacLowMacCray, TenorioTagle99}.

This we propose to be a final stage of an evolution similar in nature to
and inspired by the one presented in \citet{TenorioTagle99} and
elaborated by \citet{MasHesse03}. In this picture, initially, a
starburst is enclosed in a uniform H {\sc i} halo. As star formation proceeds,
SN feedback causes the H {\sc i} halo to expand, although unlike in the
scenario of Tenorio-Tagle, not perfectly uniformly. Growing density and
temperature perturbations in the neutral gas shell cause Rayleigh-Taylor
instabilities to arise in the expanding shell, which in time breaks up
into clumps of varying velocities, embedded in a matrix of ionized gas.
These clumps, in the LARS galaxies, have a projected covering fraction
of unity (but of course in general need not to), thus blocking all
ionizing radiation; but compared to Tenorio-Tagle's expanding shell
scenario, the overlapping clumps will allow for larger Ly\(\alpha\)
escape, and higher dust content and metal column density in the warm
neutral medium could still yield some residual intensity in the metal
absorption lines, as described in e.g.~Figure~\ref{fig:outflow}. Higher
SSFR and thus W(H\(\alpha\)) will tend to push and perforate the neutral
gas cover, while larger masses of H {\sc i} will tend to suppress this.

\section{Discussion}\label{sec:discuss}

\subsection{Comparison to 21 cm radio
observations}\label{comparison-to-21-cm-radio-observations}

A set of properties of the neutral medium of the LARS galaxies measured
by single-dish and interferometric 21 cm-band observations are presented
in \citet{LARSIII}. We shall now briefly compare this global description
of the neutral medium to our own description of the local properties
along the sight lines to the most luminous star-forming knots.

The most directly comparable quantities are the line width of the neutral medium
as measured directly, but globally, from 21 cm observations of atomic hydrogen,
and as inferred by using the low-ionization metal lines as a proxy. The numbers
are shown in Table~\ref{tab:results}, which in column 8, FWHM$_{\rm Si
\textsc{ii}}$, shows the line widths as measured from the metal lines, and in
column 9, FWHM$_{\rm H \textsc{i}}$, shows the values from \citet{LARSIII}, in
which FWHM$_{\rm H \textsc{i}}$ is reported as $W_{50}$.

\begin{figure}[htbp]
\centering
\includegraphics{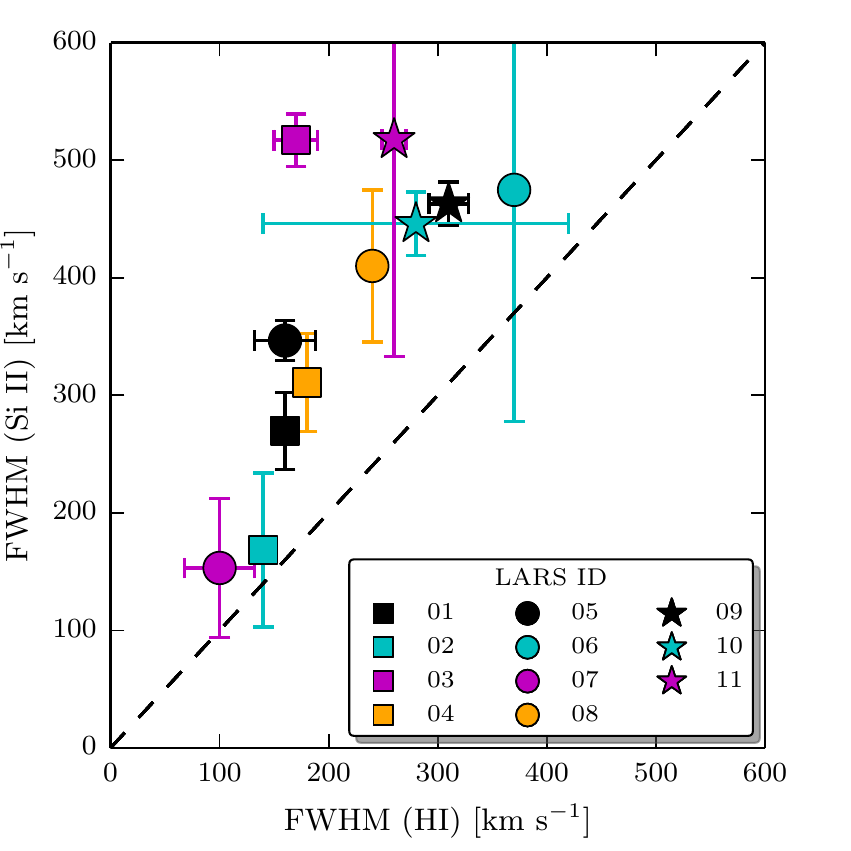}
\caption{LARS neutral medium line widths as measured from COS and GBT
spectroscopy, respectively. As expected, the measurements from the COS
aperture contain relatively more gas from regions near the hot
starbursts and therefore show wider lines than the global measurements
from the GBT. There is still a correlation; galaxies cooler globally
also tend to be cooler locally.\label{fig:fwhms}}
\end{figure}

The 21 cm measurements, which include large amounts of cool gas in the
interstellar and circumgalactic medium, show significantly lower line
widths than the COS measurements, which only target the neutral medium
along the sight lines towards the brightest and hence most energetic
star-forming knot in each galaxy. While there is no one-to-one relation,
it is evident from Figure~\ref{fig:fwhms} that there is a strong
correlation: greater metal line widths within the aperture are also
indicative of greater global line widths in H {\sc i}. The 21 cm feature is
dominated by the cool phase ISM, and its line width by large-scale
motion like e.g.~rotation. In contrast, the LIS lines are dominated by
gas in the warm neutral medium in the central regions of our galaxies,
which is heated and pushed by star formation feedback. It is therefore
not surprising that the 21 cm lines are narrower than the LIS lines; on
the other hand, given that the 21 cm lines are global and the LIS lines
such a local phenomenon, it is interesting and a bit surprising to see
them correlate so well.

\subsection{Factors governing Lyman alpha escape}\label{sec:factors}

It has been established both theoretically
\citep{Neufeldt1990, Verhamme2006} and observationally
\citep[e.g.][]{Kunth98, MasHesse03, Wofford2013} that outflows play an
important role in regulating Ly\(\alpha\) escape, with outflowing gas
being Doppler shifted out of resonance with the Ly\(\alpha\) photons
emanating from the center as the main mechanism. It is still not clear,
however, exactly how important this role is.

A number of surveys have consistently shown that stronger outflow
velocities are correlated with higher Ly\(\alpha\) escape fractions.
\citet{Kunth98} observed a sample of 8 galaxies in the local Universe
and concluded that gas kinematics, rather than dust content, is the
dominant factor governing Ly\(\alpha\) escape. A similar conclusion is
reached by \citet{MasHesse03} from a subsample of the galaxies from
\citet{Kunth98}, expands upon these results and proposes a unifying
scenario of expanding superbubbles as explanation for the different
profiles observed. In \citet{Wofford2013}, a similar conclusion about
the role of outflows is reached from a considerably larger sample of
local star-forming galaxies, with better data.

Our findings are consistent with outflows being of key importance,
although we observe this relation as an upper envelope rather than a
correlation: while stronger outflow velocity allows for a higher
Ly\(\alpha\) escape fraction, we do observe galaxies (LARS 3, 8, 11, and
13) with high wind velocities and yet very low Ly\(\alpha\) escape
fractions. These galaxies have in-aperture dust contents in the upper
half of the sample, which in its entirety has \(E(B-V)\) spanning from
\(0.047\pm 0.02\) for LARS 6 to \(0.688\pm0.015\) for LARS 3
\citep{LARSI}. In addition, their ISM span wide ranges in velocity,
further making it harder for Ly\(\alpha\) photons to be shifted out of
resonance wavelengths.

Furthermore, as is discussed in Section~\ref{sec:interp}, we suggest
that the importance of the outflowing wind lies, not only in shifting
Ly\(\alpha\) photons out of resonance, but also in creating the
Raleigh-Taylor instabilities that introduce some clumpiness and porosity
which will carve out escape routes for Ly\(\alpha\) photons through the
neutral medium. We stress again that some fragmentation of the medium is
sufficient for this to happen, no direct sight lines to the background
sources are necessary and are indeed not likely to be present in our
sample, with a possible exception in LARS 14.

For Ly\(\alpha\) photons to escape from a galaxy, they must traverse the
neutral medium, undergo a number of scatterings, for each of which it
suffers an increased chance of absorption in the meeting with a cosmic
dust grain. In a medium of zero net velocity but an internal velocity
distribution wider than the intrinsic line width of the Ly\(\alpha\)
line and with a non-zero dust content, any photons will undergo a large
number of scatterings and a resulting increase in path length to travel
out of the system. Increased path length means increased optical depth
due to dust and thus a higher probability of absorption of the photon.
It is practically impossible for radiation to escape such a system
\citep{Neufeldt1990}.

A number of factors can drastically lower the absorption probability and
thus raise the escape probability for a Ly\(\alpha\) photon:

\begin{description}
\itemsep1pt\parskip0pt\parsep0pt
\item[H {\sc i} Mass]
The mass of neutral gas in the system governs how many times a photon is
likely to scatter and therefore has a strong impact on escape
probability. Lower mass may significantly raise the escape fraction of
intrinsic Ly\(\alpha\).
\item[Neutral gas outflow]
A bulk outflow of neutral gas will Doppler shift the atomic hydrogen out of resonance with the
intrinsic Ly\(\alpha\), moving the redder parts of the line out of
resonance, from where it is free to escape.
\item[Velocity range]
A higher internal velocity range causes the hydrogen absorption profile
to broaden, requiring a larger (Doppler) redshift of the intrinsic
Ly\(\alpha\) photons to grant them easy escape. Correspondingly, a lower
velocity range will make it more likely for a photon to be redshifted
sufficiently to escape afterwards, raising the escape fraction.
\item[Dust content]
Interactions with dust grains is what destroys Ly\(\alpha\) photons, so
it is no surprise that higher dust content will result in lower escape
fraction, as is the case for any other optical and UV photons.
\item[ISM perforation]
As mentioned in Section~\ref{sec:interp}, galaxies in LARS with any
Ly\(\alpha\) escape are all drawn from the part of the sample that have
a maximum velocity-binned covering fraction
\(f_{c, \rm max} \lesssim 0.9\). This value depends strongly on the bin
width. For a bin width less than or equal to the resolution of the
spectrograph, bin width should not affect velocity-binned covering
fraction, but for larger bin widths, covering fraction will grow
monotonically with bin width as more velocities will be allowed to
contribute to each bin.
\end{description}

The interplay between these competing effects can be intricate and
unpredictable. In Figure~\ref{fig:WindEscFrac}, one can see LARS 2 far
above and LARS 3, 8, 11 and 13 falling far below what would otherwise
have been a neat correlation between outflow velocity and global
Ly\(\alpha\) escape fraction as predicted by e.g. \citet{Kunth98} and
concluded in \citet{Wofford2013}. LARS 3, 8, 11 and 13 have the highest
in-aperture dust contents of the sample \citep[table 3]{LARSI}. LARS 2,
on the other hand, has the third lowest dust content in the sample, only
surpassed by LARS 5 and LARS 6. LARS 5 has a fairly high Ly\(\alpha\)
escape fraction despite a high velocity range; the low dust content and
high outflow velocity help counter this. The medium of LARS 6 has a
strong static component, paired with the highest line width of the
sample. These conspire, despite the low dust content, to yield a very
low escape fraction. LARS 2 , on the other hand, shows the lowest
\(W_{90\%}\) of the sample. So for LARS 2, despite low outflow velocity,
dust content and low velocity range conspire to aid in a larger
Ly\(\alpha\) escape fraction, while for LARS 3, 8, 11 and 13, the
relatively high wind velocities are counteracted by a combination of
high dust content and relatively high line widths to impede Ly\(\alpha\)
escape. These factors are certainly not the only ones to govern
Ly\(\alpha\) escape, but they serve to illustrate that the problem is
multi-faceted.

\subsection{LyC leakage}\label{lyc-leakage}

The analysis of ISM covering and column density is interesting also for
the question of Lyman Continuum escape, which is important to solve the
question of which sources were responsible for reionizing the Universe.
In the local Universe, almost no leaking galaxies are observed, and the
ones that are detected show very low escape fractions. Yet we know that
around the time of reionization, the ionizing escape fraction must have
been \(\gtrsim 0.2\) \citep{Robertson2013}.

\citet{Heckman2011} analyzed a sample of 11 LBG analogs as possible
leakers of Lyman Continuum radiation. They considered two different
cases; one in which the UV sources are covered by a uniform shell with
unit covering factor, and one in which they are covered by a number of
optically thick clumps with openings between them. In the first model,
any residual flux in the line would be the result of an optically thin
medium. Since the optical depth at Lyman Continuum is significantly
larger than in the metal lines included in their analysis, such a
residual flux would be necessary but not sufficient for allowing escape
of ionizing photons.

The second model they consider is the so called ``picket fence model'',
in which the residual flux is interpreted as the result of a medium
consisting of optically thick clouds only partly covering the background
source. In a different implementation of the Apparent Optical Depth
method, they utilize the Si {\sc ii} lines present in the COS spectra to
determine which of these two cases is the more realistic. They conclude
that where residual intensity is present in their spectra, the relative
optical depths show that the picket fence model is the more plausible,
and thus conclude that galaxies in which this residual intensity is
found are likely candidates for LyC escape. They point out that these
objects all show anomalously low H\(\alpha\) emission, another pointer
that ionizing radiation is escaping. We note here that the latter
finding seems to be in contradiction to the result of this work: that a
lower \(f_C\) is quite strongly correlated with a \emph{higher}
H\(\alpha\) equivalent width. However, this might be a selection effect,
as the two samples are selected from different criteria. Observation of
a local example of a galaxy with a Picket Fence type ISM was recently
confirmed by \citet{Borthakur2014}.

\citet{Jones2013} also find that their galaxies, this time at redshifts
3 and 4, generally show residual flux in the lines, but they are more
cautious concluding that this implies any direct lines of sight (and
thus escape) through the medium. They note as a caveat that line widths
are in general much higher than the typical internal velocity dispersion
of the gas, meaning that the line is a conglomerate of contributions
from many separate subsystems. These systems do not in general occupy
the same (projected) space, and thus a low measured covering fraction is
still consistent with unit covering integrated across the line profile.
However, a lower \(f_C\) would raise the probability of the medium being
perforated to some degree, and they point out that generally, their
galaxies at \(z=4\) show lower covering than at \(z=3\). Adding the fact
that LyC escape is extremely rare in the local Universe and at low \(z\)
and extrapolating, this means that LyC escape could be even more common
earlier and thus allow galaxies to be the main source of reionization of
the early Universe.

The high resolution of the COS spectra and the accurate systemic
velocity zero points of LARS allow for thorough analysis and modeling of
the Ly\(\alpha\) line profile morphology (I. Orlitova et al., in prep.)
Visual inspection of the LARS Ly\(\alpha\) profiles shows that, with the
possible exception in LARS 2 and LARS 14 discussed above, we generally
see no signs that lower covering factors coincide with emission at line
center in Ly\(\alpha\) which is the signature of direct Ly\(\alpha\) and
LyC escape \citep[e.g.][]{Verhamme2014}. We should mention that the
optical depth in Ly\(\alpha\) is higher than in LyC, so in cases where
Ly\(\alpha\) is quenched by an attenuated interclump medium, this could
possibly still be penetrated by LyC. Only LARS 14 shows any signatures
of column densities that could possibly be low enough for this to
happen, and that shows no signs of LyC escape as inferred from the
Ly\(\alpha\) profile and the analysis of \citet{Heckman2011}.

\subsection{Scattered re-emission}\label{subs:reemission}

One possible caveat to our results is radiative transfer effects in the
metal lines of the neutral medium, in which light is reemitted into the
absorption trough of the Si {\sc ii} lines, as described by e.g.
\citet{Prochaska2011} and \citet{Scarlata2015}. The former
work shows a more general model of how this effect can work, while the
latter concerns itself more specifically with the Si {\sc ii} lines within the
COS spectral range.

In short, photons that are absorbed in the Si {\sc ii} lines will be reemitted
in random directions, meaning that there is a probability that a photon
will be scattered back in to the aperture of the spectrograph, partly
filling up the absorption trough. \citet{Prochaska2011} showed that this
effect can in certain circumstances cause a partially transparent, fully
covering system to appear as an optically thick, partly covering system
by causing transitions of a given species with different line strengths
to appear as having equal depth. Among factors that could inhibit this
effect they list a) the aperture covering only a small fraction of the
neutral medium of a galaxy, causing most of the reemission to be blocked
by the aperture, or b) the presence of fine structure splitting of the
ground state of the given transition. In the latter case, an absorbing
electron can decay into the short lived upper fine-structure level
of the ground state, at a wavelength slightly higher than the absorbing
transition, to immediately after decay into the lower fine structure
level by emitting an IR photon. The short life time of the upper fine
structure layer ensures that no absorption happens at this wavelength,
allowing a photon at this wavelength to leave the system without further
atomic scattering events.

The strength of the latter effect is difficult to estimate, as it
depends strongly on the typical number of scatterings a photon undergoes
before escaping; for each scattering, there is a fixed fractional
probability of reemission into the fluorescent channel, and the typical
number of scatterings is not well known, and could well depend strongly
on e.g.~geometry and kinematics in the system.

\citet{Scarlata2015} show in more detail how this affects the Si {\sc ii} lines
specifically. While the effect predicted in this work is more modest than the
ones predicted in Prochaska's model, they do show that it is possible with an
isotropic and fully covering outflow model to reproduce a stacked UV COS
spectrum of 25 Ly\(\alpha\) emitters, which appears to have the characteristic
Si {\sc ii} absorption signatures of an optically thick, partly covering system.
The Si {\sc ii} transitions involved here arise from a fine structure split
ground state, allowing a fraction unique for each transition to escape through
the fluorescent channel. Given the assumptions in their paper, this provides a
direct way to estimate the strength of the refilling. The resonant and
fluorescent emission happens in the same regions and under the same physical
conditions, and at a fixed ratio per scattering event. Thus, if no fluorescent
emission is present, neither will any significant resonant emission be. If
fluorescent emission is present, although there is no noticeable absorption in
the resonant channel, this can indicate that a denser neutral clump is present
inside the aperture, but off the LOS. In \citet{JaskotOey} is an interesting
discussion of how this emission, in concert with the Ly\(\alpha\) line shape,
can reveal information about the geometry of an object, in that the presence of
such fluorescent emission combined with the absence of any significant
absorption in the accompanying resonant line, reveals the presence of
significant amounts of scattering neutral gas inside the aperture but off the
LOS. Strong absorption but weak or absent fluorescent emission reveals that
scattering is present, but happening mostly outside of the aperture. The
presence of both absorption and fluorescent emission shows that absorbing gas is
present, and most of the absorbing neutral gas is found inside the aperture.
Strength and width of the Ly\(\alpha\) profile can show additional information
about the LOS column density of neutral hydrogen.

The suppression of the refilling effect due to small aperture would be
strong for some LARS galaxies, but not for all; and \citet{Scarlata2015} 
show that even with a reasonably placed and sized slit, the effect could
still be strong enough to emulate a system of opaque clumps. The
strength of suppression by escape through the fluorescent channel would
depend strongly on how many scattering events one photon would on
average undergo before escaping the system, which is not known.
Furthermore, it is unknown how strongly these effects depend on the
assumption of spherical symmetry and Sobolev approximation and how they
would react to e.g.~an anisotropic, clumpy medium.

One way to possibly assess how strong this effect would be is by looking
at the blue wing of the absorption features in question. The four Si {\sc ii}
transitions in question span more than a factor of 10 in line strength;
if they line up neatly even at wavelengths where reemission is expected
to be weak, this would suggest that the absorbing systems truly are
optically thick and only partially covering. Since reemission in the
fluorescent channel happens from the same atoms as resonant reemission,
the fluorescent lines could possibly be used to constrain whether
resonant reemission would happen in the full width of the absorption
feature and hence whether the blue line wing is clear of this
self-contamination. Preliminary measurements on the LARS galaxies
suggest that the blue wings of our absorption profiles are not behaving
differently from other parts of the absorption features.

At the current time, however, more research is needed before we know how
strong an effect, if any, this has on the line shape and the interpretation
of metal absorption lines; so at present, we shall just mention this as
a possible caveat.

\section{Summary \& Conclusions}\label{sec:conclusion}

Based on high resolution UV spectra obtained with the Cosmic Origins
Spectrograph on the Hubble Space Telescope, we have analyzed Si {\sc ii}
absorption features originating in the neutral interstellar medium along
the LOS to the strongest star-forming region of each of the galaxies in
the Lyman Alpha Reference Sample. From these lines, we have computed
kinematic properties and velocity-binned covering fractions of the
medium and compared these quantities with global properties of the
galaxies presented in \citet{LARSII}, \citet{LARSI} and \citet{LARSIII}.

We find the following:

\begin{enumerate}
\def\labelenumi{\arabic{enumi}.}
\itemsep1pt\parskip0pt\parsep0pt
\item
  Ly\(\alpha\) escape depends on several factors, including outflowing
  medium, dust content and line width. Outflows are, in our sample, a
  necessary but not sufficient condition for Ly\(\alpha\) escape. The
  conditions found to allow Ly$\alpha$ escape are generally believed to be
  markers of young age.
\item
  Of 14 galaxies, 9 show significant outflowing winds, one galaxy shows
  a generally infalling wind to more than one sigma, and four are
  consistent with a static medium. The latter show the highest
  uncertainties in outflow velocity. All galaxies with global escape
  fraction \(> 5\%\) or local emitted fraction \(> 2\%\) have outflowing
  winds.
\item
  Ly\(\alpha\) happens in sample galaxies with a neutral gas outflow
  velocity of \(v_{\rm int} \gtrsim 50 \mathrm{\, km \, s}^{-1}\),
  consistent with the findings of e.g. \citet{Wofford2013} and
  \citet{MasHesse03}. In the sample, we observe absorption-averaged wind
  velocities \(v_{\rm int}\) ranging from 68 to \(\sim\) -250 km
  \(s^{-1}\), with blue wings reaching up to a maximum of \(\sim\) -450
  km \(s^{-1}\).
\item
  Ly\(\alpha\) peak velocities for emitters in our sample do not
  correlate with the outflow speed of the wind as would otherwise be
  expected from back scattering models; in fact, we found a weak
  anticorrelation between the two. This stands in contrast to what was
  expected from backscattering models, but is consistent with the
  simpler model of an emission profile multiplied by an absorption
  feature, as seen in e.g. \citet{MasHesse03}.
\item
  Full Width at Half Maximum (FWHM) of the averaged, neutral metal lines
  correlates well with FWHM of H {\sc i} as measured in 21 cm at GBT or VLA as
  reported in \citet{LARSIII}, although the correlation is best for the
  galaxies with lower FWHM, suggesting that the galaxies of the sample
  with a neutral medium spanning larger velocity ranges also are less
  well-ordered and predictable.
\item
  Very little or no Ly\(\alpha\) escapes from galaxies with a maximum
  velocity-binned covering fraction of \(f_C \gtrsim  0.9\), although
  the \emph{total} covering fraction \(F_C\) is consistent with unity
  for all sample galaxies. Thus, a certain minimum of porosity seems to
  be necessary to allow Ly\(\alpha\) escape.
\item
  Global Ly\(\alpha\) escape is observed to be well correlated with, and
  on average around 3 times higher than, the fraction of intrinsic
  Ly\(\alpha\) emitted within the aperture.
\item
  A strong anticorrelation is found between global H\(\alpha\) EW and
  maximum velocity-binned covering fraction, suggesting that feedback
  from strong star formation helps fragmenting and perforating the
  neutral, probably expanding, medium.
\item
  We suggest an explanatory model building on that proposed by
  \citet{TenorioTagle99} and elaborated by \citet{MasHesse03}, in which
  feedback from central star-forming regions results in Raleigh-Taylor
  instabilities at the border region between the hot bubble and the
  neutral phase. These instabilities result in a clumpy neutral medium
  with hot, largely empty passageways through which Ly\(\alpha\)
  radiation can pass with significantly fewer interactions with dust
  grains than would be the case for a uniform H {\sc i} screen or spherical
  shell, resulting in a higher escape probability for each Ly\(\alpha\)
  photon.
\end{enumerate}

\section{Acknowledgments}\label{acknowledgments}
We thank the anonymous referee for a helpful suggestions leading to an
improvement in the quality of this paper.
We thank Tucker A. Jones, Claudia Scarlata and Aida Wofford for helpful
advise and useful discussions.
DK is funded by the Centre National d'Études Spatiales (CNES).  HA is supported by the European Research Council
(ERC) advanced grant ``Light on the Dark'' (LIDA) and the Centre National
d'Etudes Spatiales
(CNES). HOF was funded by a postdoctoral UNAM grant and is currently
granted by a Cátedra CONACyT para Jóvenes Investigadores. IO was
supported by the grant GACR 14-20666P of Czech Science Foundation and
the long-term institutional grant RVO:67985815. MH acknowledges the
support of the Swedish Research Council, Vetenskapsrådet, and the
Swedish National Space Board (SNSB). PL acknowledges support from the
ERC-StG grant EGGS-278202. Dark Cosmology Centre is funded by DNRF.
We acknowledge the value of high-quality scientific software made freely
available by a large community of scientists and engineers.
Specifically, This project has made extensive use of the Python-based
packages Numpy \citep{Numpy}, SciPy \citep{SciPy}, Matplotlib
\citep{Matplotlib}, Pandas \citep{Pandas}, Statsmodels
\citep{Statsmodels2010}, and others.

\bibliography{./cospaper}

\appendix
\section{Discussion of individual targets}\label{sec:targets}

\subsection{LARS 1}\label{lars-1}

LARS 1 is one of the five strong global emitters in the sample, ``strong'' being
defined by an integrated \(f_{\rm esc}^{\rm Ly\alpha} > 10\%\)
\citep[see][]{LARSII}. All four Si {\sc ii} line profiles fall inside the
detector ranges and do not seem to be polluted by other features. The strength
of the absorption differs somewhat, especially for \(\lambda 1304 \AA\), which
has the lowest resonator strength of them all, indicating that this system might
not be completely optically thick. Nonetheless, as can be seen from
Figure~\ref{fig:liproplot}, the fitted covering fractions are still close to the
absorbed fraction, indicating that complete optical thickness is still a good
approximation.

The profiles of the lines at \(\lambda \lambda\) 1193, 1304 show smaller
secondary spikes on the red side. The spike redwards of Si {\sc ii} 1193 is
emission in the fluorescent Si {\sc ii}* 1194 \(\AA\) line, which shares the
upper energy level with Si {\sc ii} \(\lambda\) 1190.

LARS 1 shows strong Lyman \(\alpha\) emission in a classic P Cygni
profile, the absorption part of which coincides neatly with the metal
absorption feature, confirming that we are tracing the same gas. While
the \(f_C(v)\) computed from the metal absorption lines shows a covering
\(< 1\) for all velocities across the line feature, the Ly\(\alpha\)
feature shows no sign of direct sight lines to the H {\sc ii} regions.

\subsection{LARS 2}\label{lars-2}

LARS 2 is another strong global Ly\(\alpha\) emitter which also locally
shows a strong emission profile in the COS aperture. Like LARS 1, it
shows a classic P Cygni-like emission profile in Ly\(\alpha\), the
absorption part of which coincides neatly in velocities with the
averaged metal line absorption profile.

The silicon absorption lines reveal a relatively narrow profile only
slightly blueshifted from the systemic velocity. The four Si {\sc ii}
transitions line up very consistently around what resembles a symmetric
Gaussian absorption feature with a shallow, secondary feature on the
blue side, which is somewhat deeper for the \(\lambda\) 1260 transition. A quite
strong fluorescent emission is present in these lines and in O \textsc{i}.

LARS 2 is peculiar because of its very high (\(= 0.56\)) global Ly
\(\alpha\) escape fraction, more than twice that of nearest competitor,
the Green Pea type galaxy LARS 14. This despite a high inferred metal
line covering fraction and low outflow velocity; factors which are
normally quenching Ly\(\alpha\) escape. The galaxy does however have
low dust content and a narrow velocity range of the neutral medium, both
of which ease the escape of Ly\(\alpha\). 

It is interesting to note that the minimum of the absorption in the
Ly\(\alpha\) profile falls bluewards of line center, leaving a
significant emission at systemic velocity. This could possibly be a
directly escaping, non-scattered component finding its way through a hole in
the medium. Such a contribution would only make up a minor fraction of
the total escaping Ly\(\alpha\), which generally shows the redshifts
typical of having undergone radiative transfer, and thus not answer the
puzzle of how such a large global escape fraction can co-exist with such
a large local covering factor.

\subsection{LARS 3}\label{lars-3}

LARS 3 is a weak Ly\(\alpha\) emitter, both globally and in the
aperture. The Ly\(\alpha\) line shows a weak P Cygni profile with a
relatively broad blue absorption component, suggesting high H
{\scshape i} column density.

All the Si {\sc ii} transitions exhibit a deep, broad absorption profile, almost
or completely saturated. The averaged profile is the widest in the sample, apart
from LARS 6, centered at \(\sim -100\) km s$^{-1}$ but with significant
absorption out to \(\sim -450\) km s$^{-1}$ on the blue side and \(\sim 150\) km
s$^{-1}$ on the red side. The strong red wing means that the O I* \(\lambda 1305
\AA\) fluorescent line stands out more conspicuously than on most other
galaxies, giving rise to the large uncertainties on \(f_C\) in the red wing
shown in Figure~\ref{fig:covfracs}, panel 3, as well as artificially slightly
raising the residual intensity in the red wing of the averaged profile. The
latter effect is counteracted by the C II \(\lambda 1334\) line being deeper
than the rest in the same velocity range.

A similar uncertainty is seen in the blue wing, probably due to the \(\lambda
1260\) line being somewhat deeper than the others here. The reason for this is
not known, as it follows the other lines very well at line center and
redwards. The effect of this on the averaged profile is likely modest, since
the 5 other lines all line up quite well in this velocity range, but it could
possibly give rise to the large uncertainties in \(f_C\) in the blue wing of the
line. The 1260 line being deeper in the blue wing is a tendency which can also
be seen in LARS 4, 9, 10, 12, 14 and less markedly so in LARS 8, leading to
believe this could be contamination from an unidentified line at \(\lambda_{\rm
rest} \sim 1259 \AA\).

\subsection{LARS 4}\label{lars-4}

LARS 4 is one of the two global Ly\(\alpha\) absorbers of the sample.
In the COS aperture, it shows weak but distinct
emission, but even in-aperture, Ly$\alpha$ absorption outweighs this.
The line shape is very similar to the one of LARS 3, although
the blue absorption component is narrower.

The averaged LIS metal line is likewise significantly narrower than that
of LARS 3, but like LARS 3 completely black at line center for all Si {\sc ii}
lines except at \(\lambda 1304\), which does show a bit of residual
intensity in the line center of LARS 4. This can partly be because this line,
having the smallest oscillator strength, is not completely opaque, but
at least part of it is likely to be due to emission in O {\sc i}*
\(\lambda 1305\) - fluorescent emission is unusually strong in this
spectrum, also for Si {\sc ii}* 1194. Along with the 1190+93 doublet, both C II
1334 , and O {\sc i} 1302 are black at line center and follow the profile of
the Si {\sc ii} doublet well. Si {\sc ii} 1260 again follows nicely in the red
wing, but is somewhat deeper in the blue wing. This could be the reason
for the discrepancy between the \(f_C(v)\) profile and the average line
profile in the blue wing; an artificially lowered residual flux for Si {\sc ii}
1260 will mean a higher inferred covering fraction due to its high value
of \(f\lambda\) while having little effect on the inferred column
density. At the same time, only four lines are included in the \(f_C\)
computations, while six are included in the average absorption profile,
lending weaker significance to one outlier.

\subsection{LARS 5}\label{lars-5}

The average LIS metal profile of LARS 5 is broad, but relatively
shallow, and shows a notable pair of minima at \(\sim -90\) km s$^{-1}$ and
\(\sim -235\) km s$^{-1}$, probably representing at least two distinct
subsystems.

Si {\sc ii} \(\lambda 1304\) is markedly shallower than the rest of the lines,
possibly indicating that the medium is not completely opaque at this
wavelength. However, the best fits at line center still agree very well
with the hypothesis of full opacity and part covering, and the
\(f_C(v)\) profile coincides remarkably well with the average line
profile. Fluorescent emission at \(\lambda \lambda 1194, 1305\) is
present but is the weakest of the galaxies so far mentioned. LARS 5 is
analyzed in depth in F. Duval et al. (in prep).

\subsection{LARS 6}\label{lars-6}

This is the second of the two global Ly\(\alpha\) absorbers, and locally
in the aperture it shows a strong, damped absorption profile.

The metal absorption profile is the most complex of the LARS galaxies.
It consists of three narrow components, one infalling at \(v \sim 280\)
km s$^{-1}$, one outflowing at \(v \sim -210\) km s$^{-1}$, and one static. The static
component is the strongest in metal absorption, but the Ly\(\alpha\)
absorption profile seems to be centered around the infalling component.

The outflowing component is interesting as the depths of the 1190+1193
doublet suggests that this component is optically thin. The transition
at \(\lambda\) 1304 follows the shape of the 1190+1193 doublet very
well, but is significantly deeper than both of these, which is unexpected,
considering that \(\lambda\) 1304 is by far the weaker of the three
lines. This is due to to the center of the infalling component of the
neighboring O {\sc i} \(\lambda\) 1302 transition coinciding with the line
center of the blue component of Si {\sc ii} \(\lambda\) 1304, so that while
looking like a single Gaussian absorption profile, it is actually the
sum of two lines.

\citet{LARSIII} show that LARS 6 is in fact interaction with
another, neighboring system. One possible explanation for the double-dip
absorption profile could be that one of the dips actually comes from a tidal
tail of gas from this interacting system extending into the aperture.

\subsection{LARS 7}\label{lars-7}

LARS 7 is another one of the strong global Ly$\alpha$ emitters and also shows
relatively strong emission within the COS aperture. The emission profile is of P
Cygni type with a narrow blue absorption component.

The strong emission in Ly\(\alpha\) is peculiar considering that the metal
absorption line, like that of LARS 2, is deep and centered close to zero
velocity, factors that should act to suppress Ly\(\alpha\) escape.

However, like for LARS 2, the velocity distribution in the neutral gas is
narrow, allowing for the red wing of Ly\(\alpha\) emission to still escape.
LARS 7 is also the faintest of the five strong global emitters, with a global
Ly\(\alpha\) escape fraction of 0.14. The in-aperture peak Ly\(\alpha\) flux for
LARS 7 is only around 1/3 that of LARS 2, while both its dust content
\citep{LARSII} and its line width are somewhat higher.

The Si {\sc ii} lines of LARS 7 show some difference in depth. The \(\lambda
1304\) feature is significantly shallower than the rest, some of which could be
fluorescent emission from O {\sc i}* 1305. The relative depths of the
$\lambda \lambda$ 1190+1193 doublet and clear emission in the corresponding
fluorescent transitions suggests that some reemission effects are present.  This
is not straightforward to interpret, but it is at least worth a word of caution
that the results of the analysis of this galaxy could be biased by these
effects.

\subsection{LARS 8}\label{lars-8}

LARS 8 is noteworthy for its strongly outflowing neutral medium. The
average metal absorption feature is black at line center, wide, and has
the highest velocity of the sample.

Despite the strong outflow, the galaxy just barely qualifies as a Ly
\(\alpha\) emitter. Globally, it has a Ly\(\alpha\) escape fraction of
\(\sim 2\%\). In the aperture, it displays a P Cygni profile weak in
emission and with a broad absorption component suggesting high H
\textsc{i} column density in the outflowing medium.

The four Si {\sc ii} lines agree well in depth, although the shape of the red
wing is a bit ambiguous; the transitions at \(\lambda \lambda 1193\) and 1260
both show a wing reaching across zero velocity and into infalling velocities,
while the two other lines show no absorption at zero velocity but do show a
minor feature on the red side, too. Some gas is present at line center, but it
does not match the shape of the Ly\(\alpha\) profile quite as neatly as many
other lines, This discrepancy could be due to a high column density of
the outflowing neutral hydrogen, meaning that the low Ly$\alpha$ flux at line
center would be due to damping wings from the P Cygni absorption feature.

Other mechanisms that could possibly account for the low
output is the fact that the neutral medium covers a large range of
velocities, thus quenching escape at more different wavelengths, and the
high dusts content of LARS 8, which is the third highest in the sample.

The computed \(f_C(v)\) show some scatter, but are generally in good agreement
with the average LIS absorption profile, supporting that the absorbing system(s)
are optically thick.  Weak fluorescent emission is apparent at O {\sc i}* 1305,
while little or no fluorescent emission seems to be present at Si {\sc ii}*
1194.

\subsection{LARS 9}\label{lars-9}

Globally a weak Ly$\alpha$ emitter, LARS 9 is the strongest absorber in the
sample measured in the COS aperture. The Ly\(\alpha\) profile is a broad,
damped absorption feature, with significant absorption out to \(\pm \sim
4000\) km s$^{-1}$, as is illustrated in Figure~\ref{fig:broadlines}.

The LIS average metal absorption profile consists of two components. A
strong, saturated static or slowly infalling component centered at
\(v \sim 50\) km s$^{-1}$, and a shallower component centered at \(\sim -220\)
km s$^{-1}$. This combination of no outflow, large velocity range and probably
high column density (judging from the depth of the metal lines)
efficiently suppresses Ly\(\alpha\) escape.

The Si {\sc ii} lines generally coincide well across the entire velocity range,
except for Si {\sc ii} \(\lambda 1260\), which is significantly deeper on the
blue side before merging with the other lines in the far blue wing. As
mentioned above, a deep \(\lambda 1260\) feature will lead to a fit that
overestimates covering fraction and underestimates column density,
which is probably what gives rise to the difference between absorption
profile and computed covering fractions on the blue side of the
\(f_C(v)\) profile.

It is unlikely that this apparent semi transparency is physical; Si {\sc ii}
1260 is the strongest of the four lines and thus the least sensitive to
changes in column density in the close to optically thick regime. Si {\sc ii}
1304 would become shallower than the rest, then the 1190+1193 doublet.
The fact that these three lines coincide so well suggests that some
other effect is influencing the blue wing of Si {\sc ii} 1260, probably a weak
contaminating line.

\subsection{LARS 10}\label{lars-10}

Much of what was written of LARS 9 above also applies to LARS 10,
although on a more moderate scale. This galaxy is also a weak emitter
globally and a strong absorber in the COS aperture, although not as
strong as LARS 9. The metal absorption lines, while black at line center
and almost static, are not as broad and not quite as strong as those of
LARS 9. Instead of a slow inflow, the line is centered at a weakly
outflowing velocity, but the line is still black at systemic velocity,
all of which efficiently blocks Ly\(\alpha\) escape.

Like for LARS 9, the Si {\sc ii} lines follow each other except for \(\lambda\)
1260, which is deeper in the blue wing than the rest. Combined with the
fact that the transitions at \(\lambda \lambda 1193\) and 1304 are
masked out in the blue wing due to contamination, this introduces some
strong uncertainty in the computations of \(f_C(v)\) bluewards of the
main line. However, this should not affect any of the main conclusions
of this work, which primarily rests on the properties at and around line
center.

\subsection{LARS 11}\label{lars-11}

LARS 11 is, along with LARS 8, notable for having a strongly outflowing
neutral medium while still having a low global Ly\(\alpha\) escape
fraction. In the COS aperture, Lyman \(\alpha\) emission is slightly
more pronounced, forming a weak P Cygni profile. As for LARS 8, the
width of the averaged LIS metal absorption profile is quite high, and
the dust content of LARS 11 is the second highest of the sample, a
combination that could possibly account for the missing Ly\(\alpha\)
output.

The Si {\sc ii} 1190+1193 doublet of LARS 11 falls between the detectors at COS
and could not be included in the analysis, meaning that the
uncertainties on inferred \(f_C(v)\) profile are large, and the
assumption of this profile coinciding with the averaged absorption
profile must in this case remain an assumption. This is further
exacerbated by the spectrum being quite noisy and the estimated
measuring errors being large. However, the Monte Carlo simulations show
that at least the estimate of the center velocity is quite robust, and
that the line persistently is completely opaque at line center.

The two Si {\sc ii} lines present both show an inflowing dip at around 250
km s$^{-1}$. However, for \(\lambda 1260\) it could well be noise, and for
\(\lambda 1304\) it might be contamination. The feature has not been
included in the analysis. This could of course have a significant
impact, especially in \(v_{int}\). However, given the large
uncertainties of the inferred values for this galaxy, these new values
would most likely be consistent with the currently given ones within
\(1 \sigma\).

None of the three Si \textsc{iv} absorption lines could be observed for
LARS 11; two fall redwards of the detector range, and one is blended with
geocoronal Ly\(\alpha\).

\subsection{LARS 12}\label{lars-12}

Globally, LARS 12 is another weak emitter, and its output in the COS
aperture is modest as well. It shows something which in
Figure~\ref{fig:coverfracs3} could look like a skewed double-peak profile,but
which, when viewing a broader range of wavelengths, turns out to be a quite
classic, weak P Cygni absorption profile.

The average LIS profile is relatively
shallow and wide. Across most of the profile, Si {\sc ii} \(\lambda 1304\)
seems slightly shallower than the rest and \(\lambda 1260\) slightly
deeper, indicating that the absorbing systems might not be completely
opaque. The best fits for \(f_C(v)\) also seem to fall a bit below the
averaged profile, although the difference is small and full opacity
still a good approximation.

The difference in line depth gets stronger on the red side, mainly due
to O {\sc i}* 1305 adding its contribution to the flux of Si {\sc ii} 1304. This
leads to uncertain and probably erratic values for \(f_C(v)\) being
reached in this region.

\subsection{LARS 13}\label{lars-13}

LARS 13 is another globally weak Ly\(\alpha\) emitter with a clear but
modest P Cygni-like emission feature inside the COS aperture. The
absorption component of this feature aligns particularly well with the
averaged LIS absorption profile as seen on Figure~\ref{fig:coverfracs3}.
The averaged LIS absorption profile is relatively wide and everywhere
well above zero, centered around \(v \sim -100\) km s$^{-1}$.

The Si {\sc ii} absorption profiles again align quite well, except
\(\lambda 1304\), which is somewhat more shallow than the rest.
Furthermore, this line and \(\lambda 1193\) both have a small peak
centered at zero velocity which is absent in the two other lines. These
two lines are the ones most strongly affected by radiative transfer
effects, but it is difficult to say exactly what creates this bump.

The transition at 1193 Å is slightly more shallow than its 1190 \AA{} counterpart,
which also indicates that some level of reemission of light in the absorption
troughs is present, although probably not much. The shallower 1304 Å line could
be due to refilling of the absorption line as well, or it could be due to the
line actually being slightly transparent at this wavelength, since this is the
weakest of these four transitions. This is reflected by the calculated values of
\(f_C(v)\) consistently being slightly higher than the values inferred from the
assumption of complete optical thickness. This effect is modest and not likely
to differ by more than 5\%, but it is difficult to say which cause is the more
important.

\subsection{LARS 14}\label{lars-14}

LARS 14 is the second strong outlier in the sample. It is a Green Pea
type galaxy \citep{GreenPea}, a class of compact, hot galaxies with
low neutral gas column density. The Ly\(\alpha\) profile of LARS 14
shows strong emission in a double-peaked profile, the central absorption
feature of which is remarkably shallow, with a minimum significantly
above continuum level, indicating a low H {\sc i} column density. This is
further supported by the Si {\sc iv} absorption profile being significantly
deeper than the averaged LIS metal line, indicating an ionization degree
unusually high for the sample.

The average LIS metal absorption profile consists of two shallow and
narrow components. One is centered at \(v\) close to zero and just
barely significant, but persistent across all six LIS lines. The other
is somewhat stronger but still very shallow, with a minimum residual
intensity of around 40\%. It is centered at \(v \sim 320\) km s$^{-1}$. The
central component sees its Si {\sc ii} profiles lining up very well, although
this, given the weakness of the feature and the measurement
uncertainties, should be interpreted with caution. However, if this is
in fact true, it indicates that the neutral medium at these velocities
is optically thick but fragmented. The stronger, outflowing component
shows a somewhat deeper feature at \(\lambda 1260 \AA\) and a somewhat
shallower feature at \(\lambda 1304 \AA\), consistent with a medium that
is not fully opaque at these wavelengths, as is also apparent from the
\(f_C(v)\) profile in Figure~\ref{fig:coverfracs3}. This should,
however, also be interpreted with caution, due to the uncertainties
involved.

The uncertainties combined with the two shallow components of the
profile also give a high uncertainty in the calculated dynamic
velocities: when performing the Monte Carlo simulation, there is a
relatively high probability that the uncertainties drawn can line up to
shift the weight of the profile from one component to the other.

The Si {\sc ii} profiles are peculiar in that they show very modest
absorption/scattering along the line of sight, but still strong
fluorescent emission in O {\sc i}* 1305 and Si {\sc ii}* 1194. We suggest the
explanation that significant scattering and re-emission in a relatively
dense neutral medium is happening inside the aperture, but off the line
of sight to the central sources, as described in \citet{JaskotOey}. We
interpret this as further indication of a clumpy, fragmented neutral
medium.

Green Peas have been suggested as candidates for Lyman Continuum leaking
galaxies  \citep[e.g.][]{Jaskot2013, Nakajima2014} due to their very H {\sc i}
column density. This would, in the case of LARS 14, be supported by the
overall low values of \(f_C(v)\), with a maximum value of 0.4. However,
\citet{Heckman2011} ruled out that this galaxy should leak ionizing
radiation along our line of sight, and we find no evidence against this
claim.

\end{document}